\documentclass[a4paper,11pt]{article}
\pdfoutput=1

\usepackage{subcaption}
\usepackage{placeins}
\usepackage{gensymb}

\usepackage{jinstpub}
\usepackage{lineno}

\usepackage{orcidlink}
\hypersetup{colorlinks=true,linkcolor=black,citecolor=black,urlcolor=black,filecolor=black}

\title{Performance of a first multi-cell WOM-based liquid scintillator detector as prototype for the SHiP Surrounding Background Tagger}

\author[a]{M.~B\"ohles\orcidlink{0000-0002-8486-8557},}
\author[b,*]{A.~Brignoli\orcidlink{0009-0001-4190-7026},}
\author[a]{P.~Deucher\orcidlink{0000-0003-2793-4666},}
\author[b]{C.~Eckardt\orcidlink{0009-0002-0342-4899},}
\author[c]{H.~Fischer\orcidlink{0000-0002-9342-7665},}
\author[a,f,*]{A.~Hollnagel\orcidlink{0000-0003-0040-8420},}
\author[c]{A.~Krolla\orcidlink{0009-0009-3135-5350},} 
\author[b,*]{H.~Lacker\orcidlink{0000-0002-7183-8607},}
\author[c]{P.~Luther\orcidlink{0009-0001-2320-5231},}
\author[c,*]{F.~Lyons\orcidlink{0009-0005-6444-4422},}
\author[a]{J.~Molins~i~Bertram\orcidlink{0009-0000-5377-3312},}
\author[c]{T.~Molzberger\orcidlink{0009-0009-4616-6632},}
\author[c]{A.~S.~Müller\orcidlink{0009-0003-1270-1859},}
\author[c]{S.~Ochoa~Guaman\orcidlink{0000-0003-1423-5989},}
\author[d]{F.~Rehbein\orcidlink{0009-0005-1713-5432},}
\author[c]{T.~J.~Rock\orcidlink{0009-0005-9464-2024},}
\author[e]{M.~Schaaf\orcidlink{0000-0001-5148-6627},}
\author[b]{C.~Scharf\orcidlink{0000-0002-0294-1205},}
\author[c]{M.~Schumann\orcidlink{0000-0002-5036-1256},}
\author[c]{J.~M.~Webb\orcidlink{0000-0002-5294-6856},}
\author[c]{J.~Wenk\orcidlink{0009-0006-2067-7950},}
\author[b]{I.~Wöstheinrich\orcidlink{0009-0005-8083-9176},}
\author[a]{M.~Wurm\orcidlink{0000-0003-2711-0915}}

\affiliation[a]{Institut für Physik \& Exzellenzcluster PRISMA$^+$, Johannes Gutenberg-Universität Mainz, Staudingerweg 7, 55128 Mainz, Germany}
\affiliation[b]{Institut für Physik, Humboldt-Universität zu Berlin, Newtonstraße 15, 12489 Berlin, Germany}
\affiliation[c]{Physikalisches Institut, Albert-Ludwigs-Universität Freiburg, Hermann-Herder-Straße 3, 79104 Freiburg, Germany}
\affiliation[d]{Physikalisches Institut 3A, Rheinisch-Westfälische Technische Hochschule Aachen, Otto-Blumenthal-Straße 16, 52074 Aachen, Germany}
\affiliation[e]{Zentralinstitut für Engineering, Elektronik und Analytik -- Engineering und Technologie, Forschungszentrum Jülich GmbH, Wilhelm-Johnen-Straße, 52428 Jülich, Germany}
\affiliation[f]{Present address: International Center for Hadron Astrophysics, Chiba University, 1-33 Yayoi-cho, Inage-ku, 263-8522 Chiba-Shi, Japan}
\affiliation[*]{Corresponding authors}

\emailAdd{brignoli@physik.hu-berlin.de}
\emailAdd{annika.hollnagel@uni-mainz.de}
\emailAdd{lacker@physik.hu-berlin.de}
\emailAdd{fairhurst.lyons@physik.uni-freiburg.de}

\date{July 2026}

\abstract{
    The Search for Hidden Particles (SHiP) Experiment was approved by CERN in~2024. Feebly-interacting particles that are produced in a proton Beam Dump Facility (BDF) will decay in the 50\,m-long Decay Volume of the experiment, which needs to be enveloped by a hermetic veto detector: The Surrounding Background Tagger (SBT). Its technology relies on liquid scintillator, composed of linear alkylbenzene and 2,5-diphenyloxazole, as active detector material and Wavelength-shifting Optical Module (WOM) tubes collecting the primary scintillation photons. The liquid scintillator volume is segmented in large cells of typically \hbox{$120\,\mathrm{cm}\,\times\,80\,\mathrm{cm}\,\times\,20\,\mathrm{cm}$} that are equipped with two WOMs each. Here, we report on the performance of a full-scale $2\times2$-cell prototype detector which was exposed to 5\,GeV muons at the CERN~PS~T9 test beam facility to study the detector response and its time and spatial resolution for minimum ionising particles crossing multiple detector cells.}

\begin{document}
\maketitle
\flushbottom

\section{Introduction}
\label{Sec:Introduction}

The Search for Hidden Particles (SHiP) experiment has recently been approved to be built at the future SPS proton Beam Dump Facility~(BDF)~\cite{SHiP:2015vad,SHIP:2021tpn} in CERN's North Area experimental cavern ECN3~\cite{SHiP-LoI}. To reduce the number of background reactions induced by neutrinos and muons, the large Decay Volume of the SHiP Hidden Sector~(HS) detector will be filled with helium at atmospheric pressure. This Decay Volume has to be enveloped by the Surrounding Background Tagger~(SBT) to enable the tagging of muons in the energy range of $1\,\mathrm{GeV} - 400\,\mathrm{GeV}$ that enter the Decay Volume from the outside, as well as inelastic interactions of muons and neutrinos with the helium inside the Decay Volume, the SBT, and its vicinity. A segmented metal structure of cells filled with liquid scintillator~(LS) as active detector material will provide optimal hermeticity for this veto tagger, as well as high detection probability.

The SBT must be able to efficiently detect the energy depositions of minimum ionising particles~(MIPs) crossing its 20\,cm-thick liquid scintillator layer at minimal path length. Background rejection studies of muon or neutrino inelastic interactions in SHiP have assumed a benchmark detection efficiency of more than 99\%~\cite{SHiP-LoI}. Further requirements for the SBT are a time resolution in the nanosecond range and a spatial resolution of better than 20\,cm. This will allow the identification of muons entering the Decay Volume from the outside, and to distinguish the decays of feebly-interacting particles inside the Decay Volume from particles produced by inelastic scattering of neutrinos or muons off the helium, which hit the SBT from the inside~\cite{SHiP-LoI,SHiP-Proposal}.
The fulfilment of these performance benchmarks has already been demonstrated with several prototype detector cells made from different materials~\cite{Ehlert:2018pke,Alt:2023vuu,Brignoli:2025spc}.

A dangerous background for SHiP is posed by muons produced in the target of the BDF. These will enter the detector at very shallow angles, thus passing more than one SBT detector~cell. To study the detector response in cases where at least two SBT cells are hit, we constructed a multi-cell detector prototype consisting of 2\,$\times$\,2 LS cells of about \hbox{120\,cm\,$\times$\,80\,cm\,$\times$\,20\,cm}, similar to the size of previous single-cell detectors~\cite{Alt:2023vuu}.

\FloatBarrier
\section{Design and construction of the multi-cell detector prototype}
\label{Sec:LiquidScintillatorDetectorCell}

\begin{figure}[hbt]
    \centering
    \includegraphics[width=\textwidth]{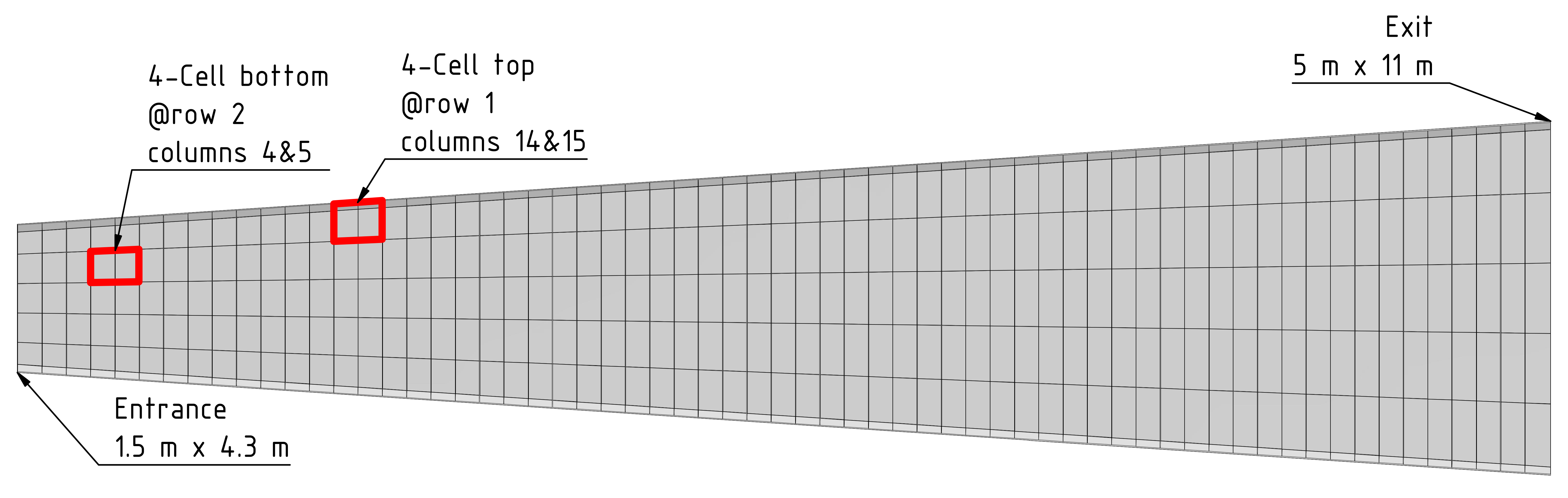}
    \caption{SBT detector design, status 2021~\cite{SHIP:2021tpn}. The positions of the exemplary detector cells that make up the four-cell prototype studied in this publication are highlighted in red.}
    \label{fig:sbt2024}
\end{figure}

\begin{figure}[htb]
    \centering
    \begin{subfigure}[c]{0.6\textwidth}
        \includegraphics[height=6.6cm]{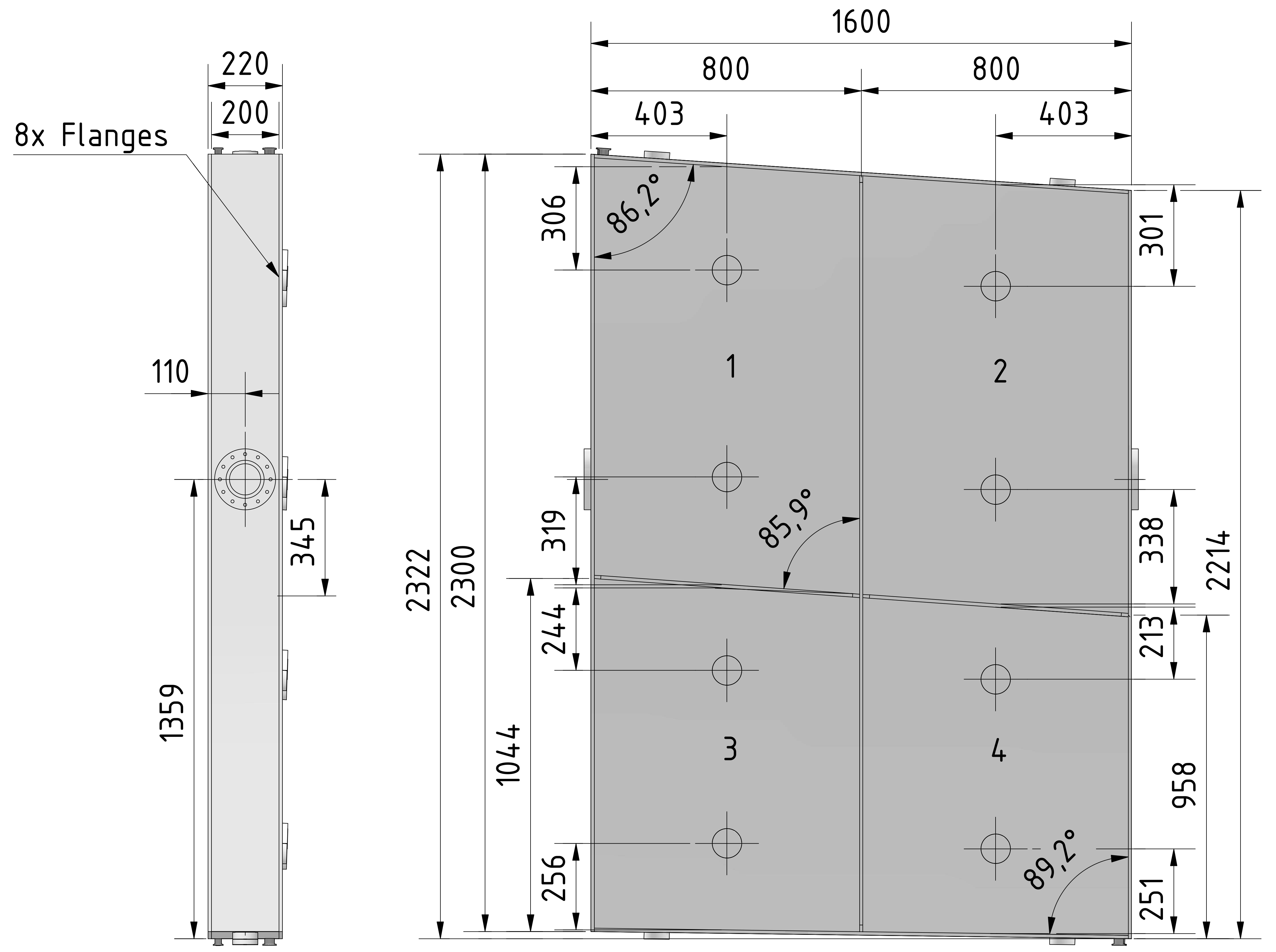}
        \vspace{4.5mm}]
        \caption{Prototype CAD drawing.}
        \label{fig:cella}
    \end{subfigure}
    \hspace{10mm}
    \begin{subfigure}[c]{0.3\textwidth}
        \vspace{5mm}
        \includegraphics[height=6.6cm]{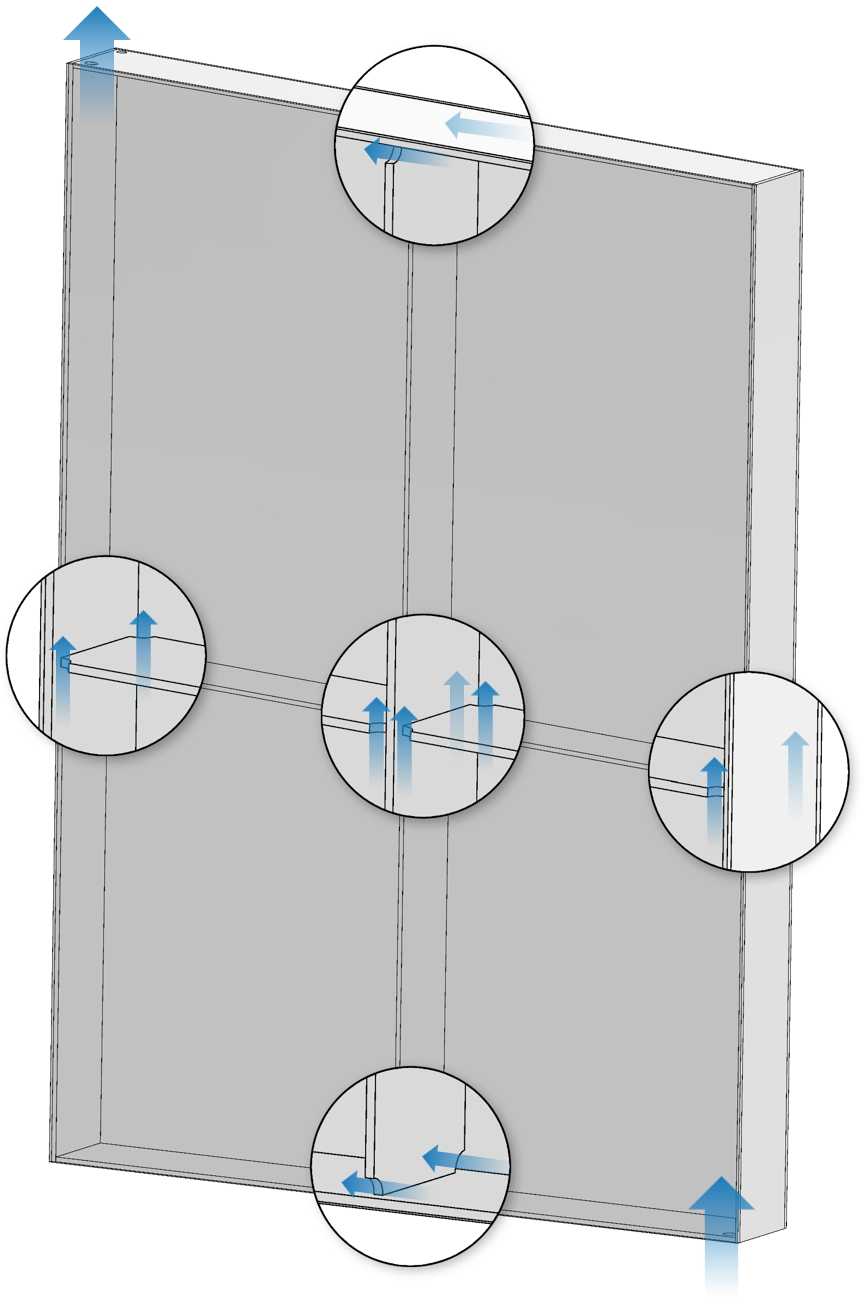}
        \caption{Liquid scintillator flow.}
        \label{fig:ls4cell}
    \end{subfigure}
    \caption{(a) CAD drawing of the four-cell SBT detector prototype. All dimensions are given in mm, the thickness of the COR-TEN\textsuperscript{\textregistered} steel walls is 10\,mm. The cells are labelled by numbers. (b) Liquid scintillator flow inside the interconnected cells of the prototype detector during filling.}
    \label{fig:cell}
\end{figure}

This work describes a dedicated test beam campaign for the characterisation of the performance of a multi-cell prototype detector that is representative of a part of the SBT (see Figure~\ref{fig:sbt2024}). As was originally foreseen in the context of an evacuated Decay Vessel~\cite{SHIP:2021tpn}, this detector prototype was constructed from COR-TEN\textsuperscript{\textregistered} steel sheets (S355~J2W~1.8965) of 10\,mm thickness that were welded together. The resulting four detector cells had an inner width of 785\,mm and a height varying from 938\,mm to 1246\,mm (see Figure~\ref{fig:cella}), corresponding to the widening frustum shape of the Decay Volume. With an inner LS layer thickness of 200\,mm, the outer cell depth was 220\,mm. The total weight of the filled detector prototype amounted to about 1\,800\,kg.

The light produced in the liquid scintillator was collected by Wavelength-shifting Optical Module~(WOM) tubes~\cite{Bastian-Querner:2021uqv}. To accommodate the PMMA vessels housing the WOMs, each detector cell had two circular openings of 70\,mm diameter on its front plate (horizontally centred and vertically positioned half-way between cell wall and cell centre) that were equipped with \hbox{COR-TEN\textsuperscript{\textregistered}} steel flanges. 

At the highest and lowest points of the multi-cell detector, two short stainless steel ISO-KF25 flanges enabled filling and emptying of the LS cells: The bottom opening could be closed with a ball valve, while the top opening was connected to an expansion vessel ensuring complete detector filling despite possible expansion / contraction of the liquid volume with changing temperature. Additional ISO-KF16 connections at the top and bottom were used for pressure monitoring and liquid level control. Fluid exchange between the four detector cells was enabled via of 20\,mm-radius quarter-circle openings in the corners of the separating walls, internally connecting the cells (see~Figure~\ref{fig:ls4cell}). The detector was mounted on shaft axes via two blind flanges that were fitted to its sides, and two lifting lugs welded to each the top and the bottom of the detector allowed for transport. 

After welding, the inside of the detector cells was spray-coated via the WOM flange openings with two layers of primer followed by two layers of diffuse-reflective Optopolymer\textsuperscript{\textregistered} BaSO$_4$ paint (Berghof Fluoroplastics Technology GmbH)~\cite{deucher2026,baso4}. This significantly increased the surface reflectivity and consequently the light-collection efficiency of the detector. To mitigate the development of rust stains that had locally lowered the surface reflectivity of the previous one-cell detector prototype~\cite{Alt:2023vuu} likewise constructed from COR-TEN\textsuperscript{\textregistered} steel, a dedicated anti-rust primer was used.

The detector was completely filled with LS made from alumina-purified linear alkylbenzene~(LAB) as solvent and 2\,g/l of \hbox{2,5-diphenyloxazole~(PPO)} as fluor. $\alpha$-Tocopherol was added at a concentration of 0.001\% as an oxidation inhibitor.

Each cell was equipped with two WOMs made from PMMA tubes with a length of 205\,mm, an outer diameter of 60\,mm, and a wall thickness of 3\,mm. The WOM tubes were dip-coated on both the outside and the inside with a dye made from toluene and Paraloid~B723~(PEMA) as base material and bis-MSB and p-terphenyl~(PTP) as wavelength shifters, at a ratio of 72.83\,:\,26.69\,:\,0.156\,:\,0.324.
One end of each WOM tube was instrumented with a circular printed circuit board (PCB) holding an array of 40 silicon photomultipliers~(SiPM). A detailed discussion of the readout will be given in Section~\ref{Sec:ElectronicsAndDAQ}.

\FloatBarrier
\section{CERN test beam results}
\label{Sec:CERNtestbeam}

In order to test its performance, the prototype detector was exposed to a 5\,GeV muon beam at the CERN PS East Area T9 between March and April 2024.

\FloatBarrier
\subsection{Measurement setup}
\label{Sec:DetectorSetup}

A complete characterisation of the four-cell detector prototype required the measurement of signals produced by incoming particles from multiple angles and covering the entire detector volume. This was achieved by a positioning system with four degrees of freedom, enabling the realisation of all possible detector orientations relative to the horizontal muon beam at fixed height. Vertical and horizontal perpendicular translation was achieved by means of a crane and rail system, and the detector could be rotated around its vertical and horizontal axes. The origin of the coordinate system was defined by the intersection of the inner detector walls separating the four cells. 

\begin{figure}[ht]
    \centering
    \begin{subfigure}[c]{0.53\textwidth}
        \includegraphics[height=17.0cm]{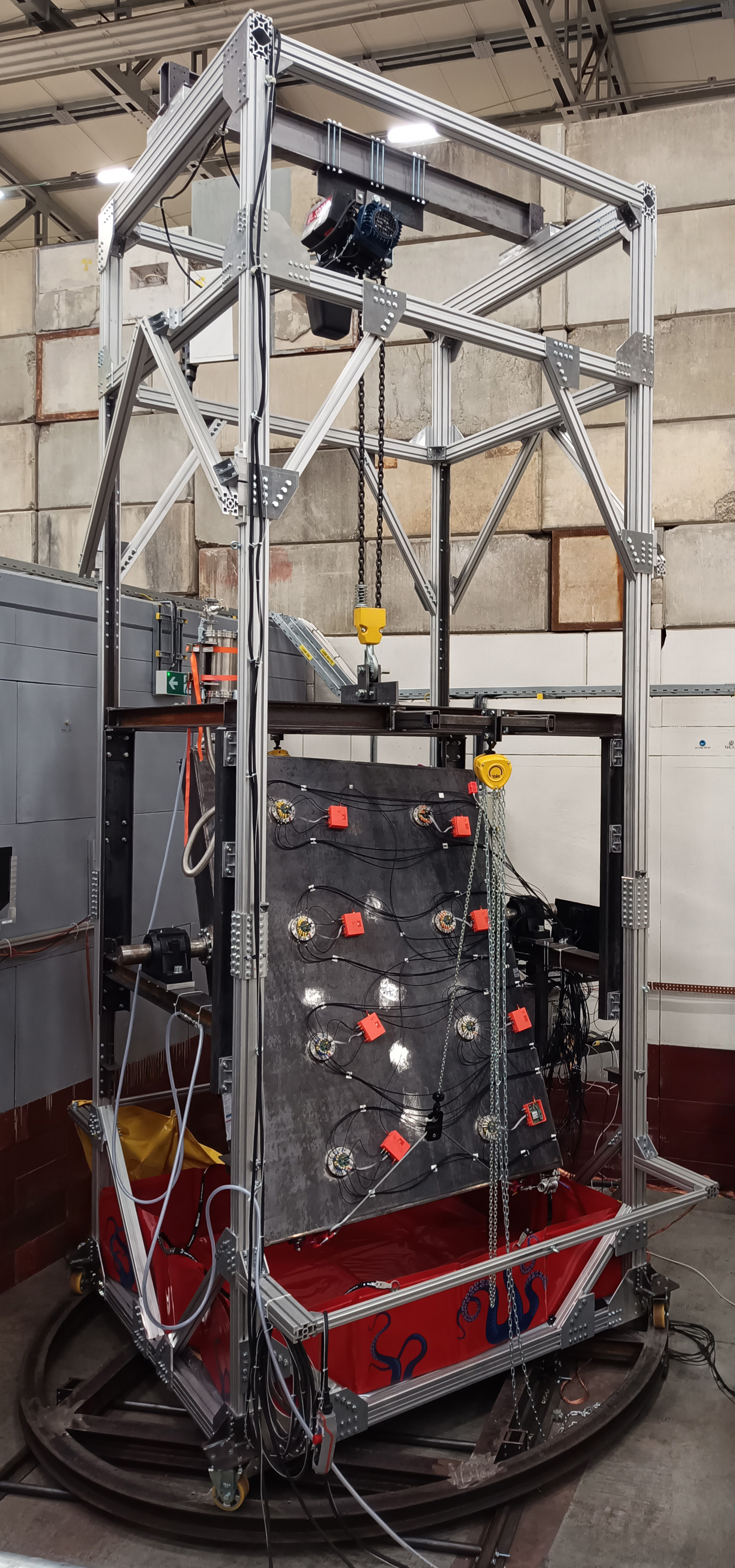}
        \caption{Detector in Holding Tower.}
        \label{fig:setup}
    \end{subfigure}
    \hspace{5mm}
    \begin{subfigure}[c]{0.42\textwidth}
        \centering
        \includegraphics[height=8.35cm]{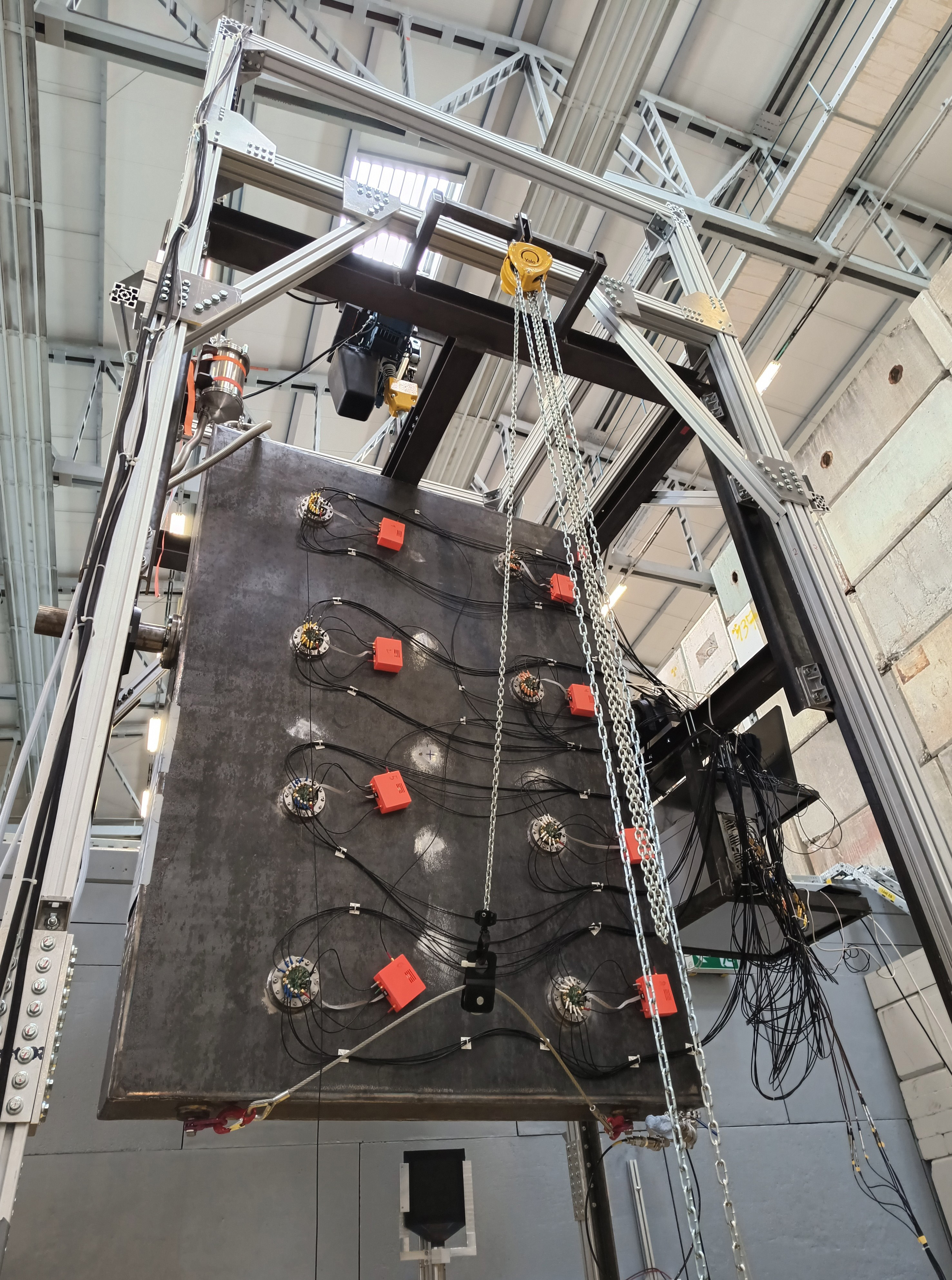} \\
        \caption{Elevation via crane.}
        \label{fig:cellframe}
        \vspace{5.0mm}
        \includegraphics[height=7.45cm]{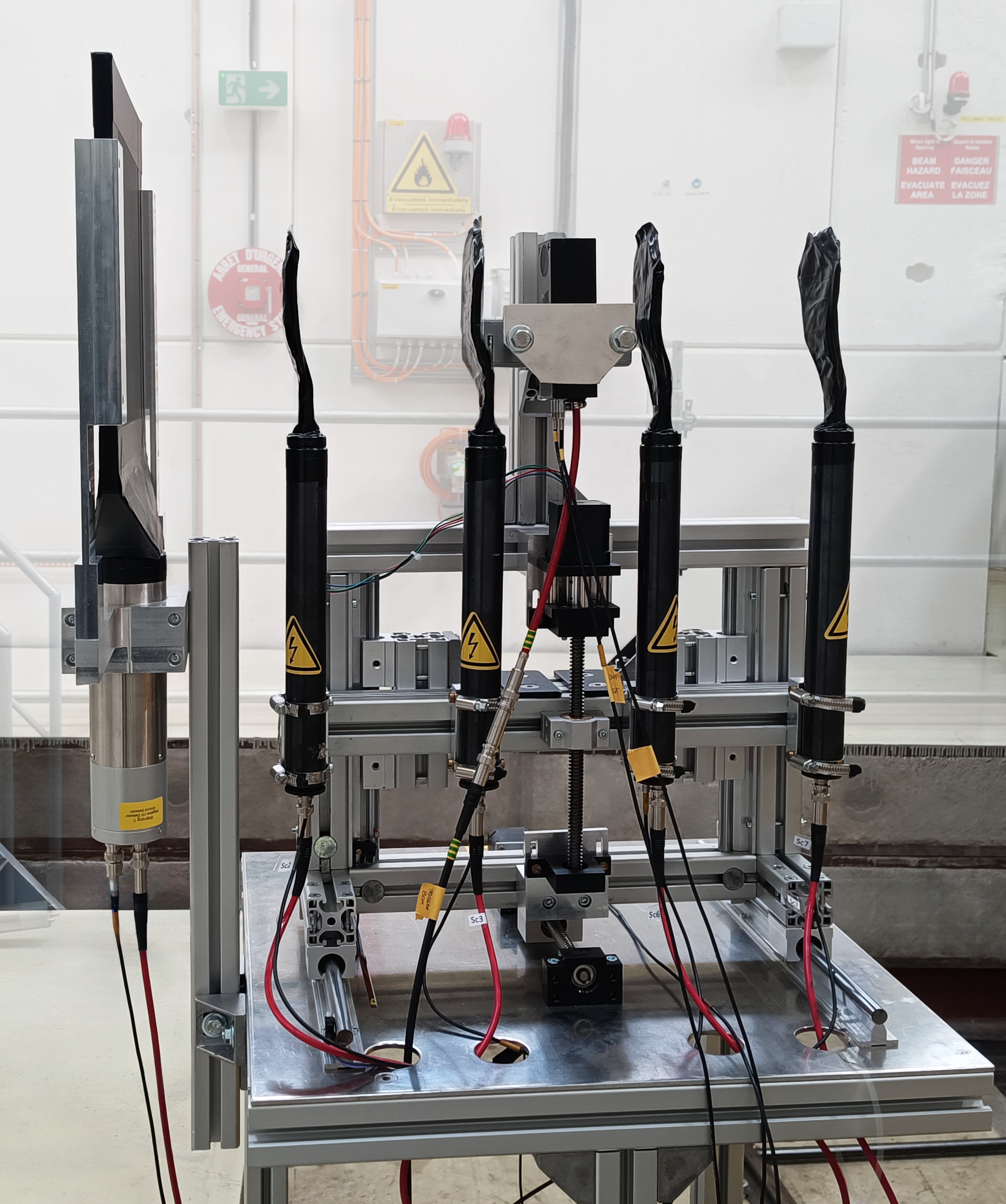}
        \caption{Beam Telescope.}
        \label{fig:beamtelescope}
    \end{subfigure}
    \caption{(a) The four-cell detector prototype mounted in its Holding Tower offering four degrees of freedom. Positive rotation in $\theta_X$ tilts the detector front plate (equipped with WOMs) facing up, rotation in $\theta_Y$ occurs anti-clockwise. $\theta_X = 0\degree$ and $\theta_Y = 0\degree$ are defined by the beam perpendicularly crossing the detector front and back walls. (b) Front view of the detector prototype at $\theta_X = 0\degree$ and $\theta_Y = 0\degree$ while elevated in the Holding Tower. (c) Beam Telescope made of four thin plastic scintillators of 5\,cm diameter and a thin plastic scintillator of 2\,cm diameter sandwiched in the centre. The beam enters the telescope from the left, through a large plastic scintillator panel.}
    \label{fig:holding-structure}
\end{figure}

Figure~\ref{fig:setup} and~\ref{fig:cellframe} show the detector Holding Tower. It was built from aluminium and steel profiles, with an area of 3.2\,m\,$\times$\,3.2\,m, a vertical height of 6.2\,m, a horizontal rail length of 6\,m, and a weight of 2.5\,t. The structure offered a maximum load capacity of 2.5\,t, with 3\,m horizontal moving range ($X$), 1.85\,m vertical moving range ($Y$), full 360$\degree$ rotation around the vertical axis ($\theta_Y$) and $\pm 90\degree$ tilt around the horizontal axis ($\theta_X$). Vertical movement was controlled electrically, while all other movements were carried out manually. 

Upstream of the detector, a Beam Telescope consisting of four $3.1$\,mm-thick plastic scintillators of 5\,cm diameter and a 1\,mm-thick plastic scintillator of 2\,cm diameter in the centre (see Figure~\ref{fig:beamtelescope}) was installed to increase the location accuracy of incoming beam particles. Two additional larger panels of plastic scintillator with a thickness of 1.0\,cm and an area of 28\,cm\,$\times$\,28\,cm were placed upstream and downstream of the detector for identifying particles traversing the four-cell prototype and determining its detection efficiency.

\FloatBarrier
\subsection{Readout electronics and data acquisition}
\label{Sec:ElectronicsAndDAQ}

During the R\&D phase of the SHiP SBT, the detection and subsequent processing and readout of the light signals produced in the prototype detectors relies on a triad of hardware: A custom-designed SiPM carrier board, an eight‑channel eMUSIC~\cite{MUSIC-ASIC} pre‑amplifier board, and a WaveCatcher~\cite{WaveCatcher} digitiser. 

At its cell flange side, each WOM was coupled with optical grease to a ring‑shaped array of 40~Hamamatsu \hbox{S14160‑3050HS} SiPMs with an effective photosensitive area of 3\,mm\,$\times$\,3\,mm that were operated at an overvoltage of~$\sim$3.7\,V~\cite{Hamamatsu-S14160}. The SiPM board, which is composed of a 2.08\,mm-thick four-layer FR-4 substrate, provides a common bias voltage and routes the analogue outputs to a high-density, 100-pin SAMTEC connector. The bias voltage is stabilised by a low-noise CAEN A7585D regulator and blocked locally with 10\,nF capacitors at each SiPM cathode. All vias are filled to prevent light penetrating from the outside to the photodetectors. An on‑board temperature-sensing integrated circuit \hbox{(TMP37GRTZ-REEL7)} provides real‑time monitoring for gain - temperature corrections.

On the eMUSIC pre-amplifier board, the~40 analogue SiPM signals are summed into eight groups per five adjacent SiPMs. These are then fed into an eight-channel eMUSIC-8 ASIC that provides a programmable gain with either 180\,$\Omega$ or 480\,$\Omega$ for its transimpedance amplifier, followed by a shaper and a pole‑zero cancellation network. Configuration of the ASIC registers is performed via SPI using a low‑power Raspberry Pi Pico~W microcontroller~\cite{raspberry}, which also monitors the bias voltage distribution with on‑board decoupling capacitors. The amplified signals are then provided on SMA connectors with 50\,$\Omega$ termination for transmission to the digitiser. Throughout all measurements in this test beam campaign, the high gain setting of the eMUSIC ASIC has been used.

Digitisation of the signal waveforms is carried out by a switched‑capacitor‑array 64-channel WaveCatcher module with a selectable sampling rate up to 3.2\,GS/s, an analogue bandwidth of 500\,MHz, and 4\,096\,samples per trigger and channel~\cite{WaveCatcher}.

This combination of SiPM carrier board, eMUSIC frontend, and WaveCatcher digitiser creates a low‑noise readout solution with high dynamic range that enables both precise charge measurement and sub‑nanosecond timing for the WOM‑based liquid scintillator detector cells of the SHiP SBT.

Data acquisition by the WaveCatcher was triggered via a coincidence of three out of the four 5\,cm diameter Beam Telescope scintillators (see Figure \ref{fig:beamtelescope}) and the large plastic scintillator panel downstream of the setup. While the small scintillator at the centre of the Beam Telescope was not used the trigger, it was included in later data analysis steps by providing time signals and reducing the effective beam spot size on the detector.

\FloatBarrier
\subsection{Collected data}
\label{Sec:CollectedData}

\begin{figure}[htb]
    \centering   
    \includegraphics[width=0.6\textwidth]{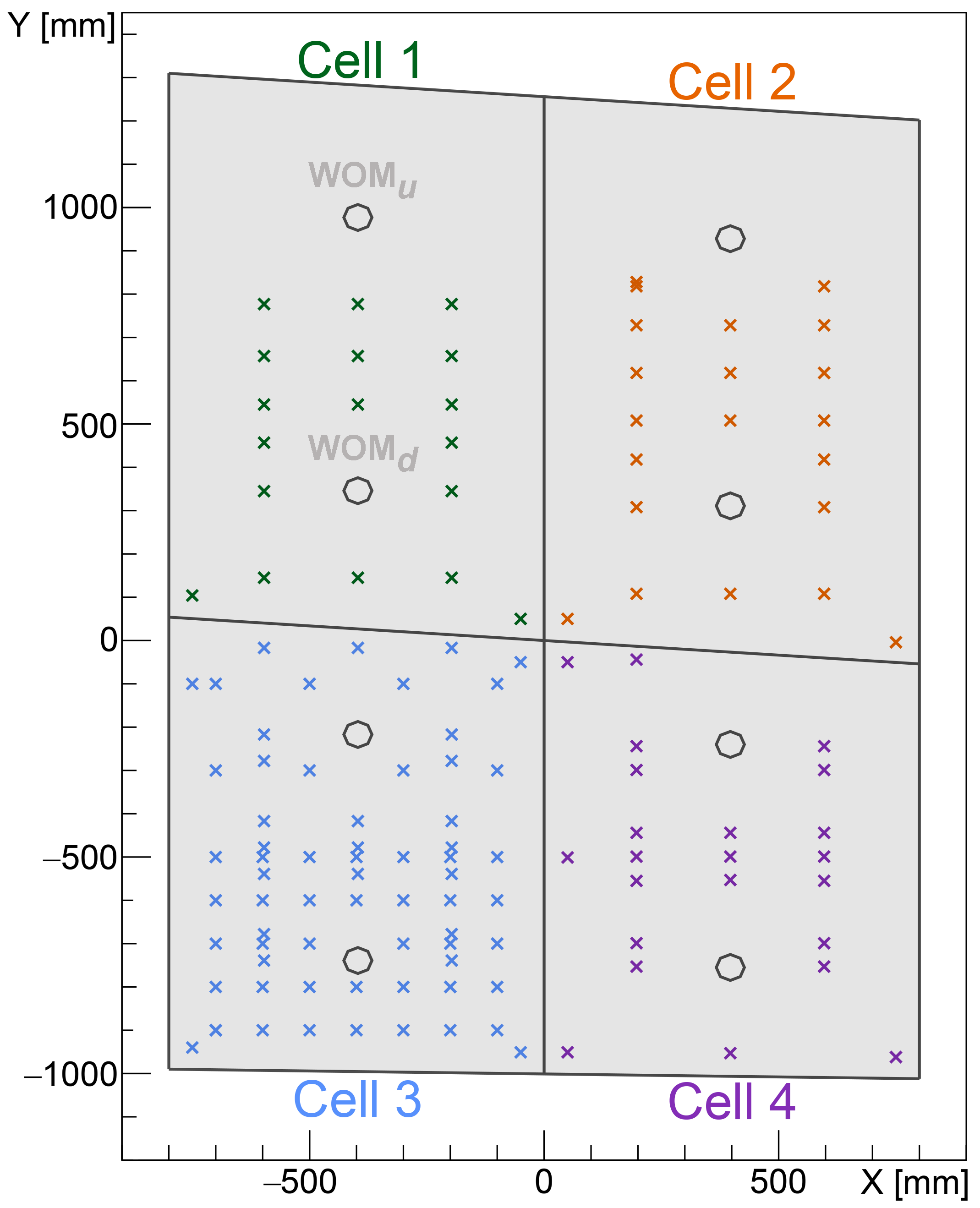}
    \caption{Locations of particle crossing points for a beam incident angle of $\theta_X=0\degree$ and $\theta_Y=0\degree$ on the detector front plate, also illustrating the coordinate system used throughout this publication: The origin at $X=0$\,mm and  $Y=0$\,mm is defined by the intersection of the inner detector walls separating the four detector cells, $Z=0$\,mm describes a central plane through the detector. With a LS thickness of $200$\,mm, the front plate with the WOM flanges is at $Z=100$\,mm, and the detector back plate at $Z=-100$\,mm. Cell walls and WOM locations are indicated by outlines, crosses mark the locations where the particle beam traversed the central plane of the detector during the measurements.}
    \label{fig:deglocations}
\end{figure}

A large number of particle crossing points on the detector cells were measured, also realising various angles between detector and muon beam. The measured particle crossing points at an incident angle of $\theta_X=0\degree$ and $\theta_Y=0\degree$ between the beam and the normal to the detector front plate are shown in Figure~\ref{fig:deglocations}. In~the following, the WOMs of each cell at larger $Y$ coordinates will be referred to as WOM$_{u}$ (\textit{up}) and the ones at smaller $Y$ coordinates as WOM$_{d}$~(\textit{down}). Per~run, at least 5\,000~events were recorded. 

\FloatBarrier
\subsection{Detector response}
\label{Sec:DetectorResponseLightCollection}

\FloatBarrier
\subsubsection{Light collection and detector response uniformity}
\label{Sec:LightCollectionandDetectorResponseUniformity}

The integrated light yield \textit{LY} describes the amount of light detected per event and is quantified by integrating the SiPM waveforms over time, in units of V\,$\times$\,ns: The integrated light yield per channel will be called $LY_\text{ch}$, the sum of integrated light yield for a WOM $LY_\text{WOM}$, and the sum of integrated light yield for the entire cell $LY_\text{cell}$. In the absence of explicit gain calibrations for each SiPM, this approach provides an alternative to calculating the total number of detected photons. The length of the integration time window for each waveform is chosen to be 120\,ns: 20\,ns before the peak maximum, and 100\,ns after. This retains most information of the waveform and at the same time keeps the amount of dark counts at a sufficiently low level. A baseline correction of the waveform is conducted using the minimum root-mean-square slope of the waveform within a sliding time window of 25\,bins (25\,$\times$\,0.3125\,ns) inside a 50\,ns-window before the start of the signal. Events are rejected if $LY_\text{WOM}$ is lower than 0.8\,V\,$\times$\,ns, defined by the largest random trigger background signal measured by a WOM when the beam is turned off (usually corresponding to dark counts, see \cite{Alt:2023vuu} for details). To reduce the contributions of hadrons from beam contamination creating hadronic showers in the detector, also events with an integrated light yield $LY_\text{WOM}$ exceeding 50\,V\,$\times$\,ns in one WOM are rejected for perpendicular tracks ($\theta_X=0\degree$ and $\theta_Y=0\degree$).

\begin{figure}[htb]
    \centering
    \includegraphics[width=0.85\textwidth]{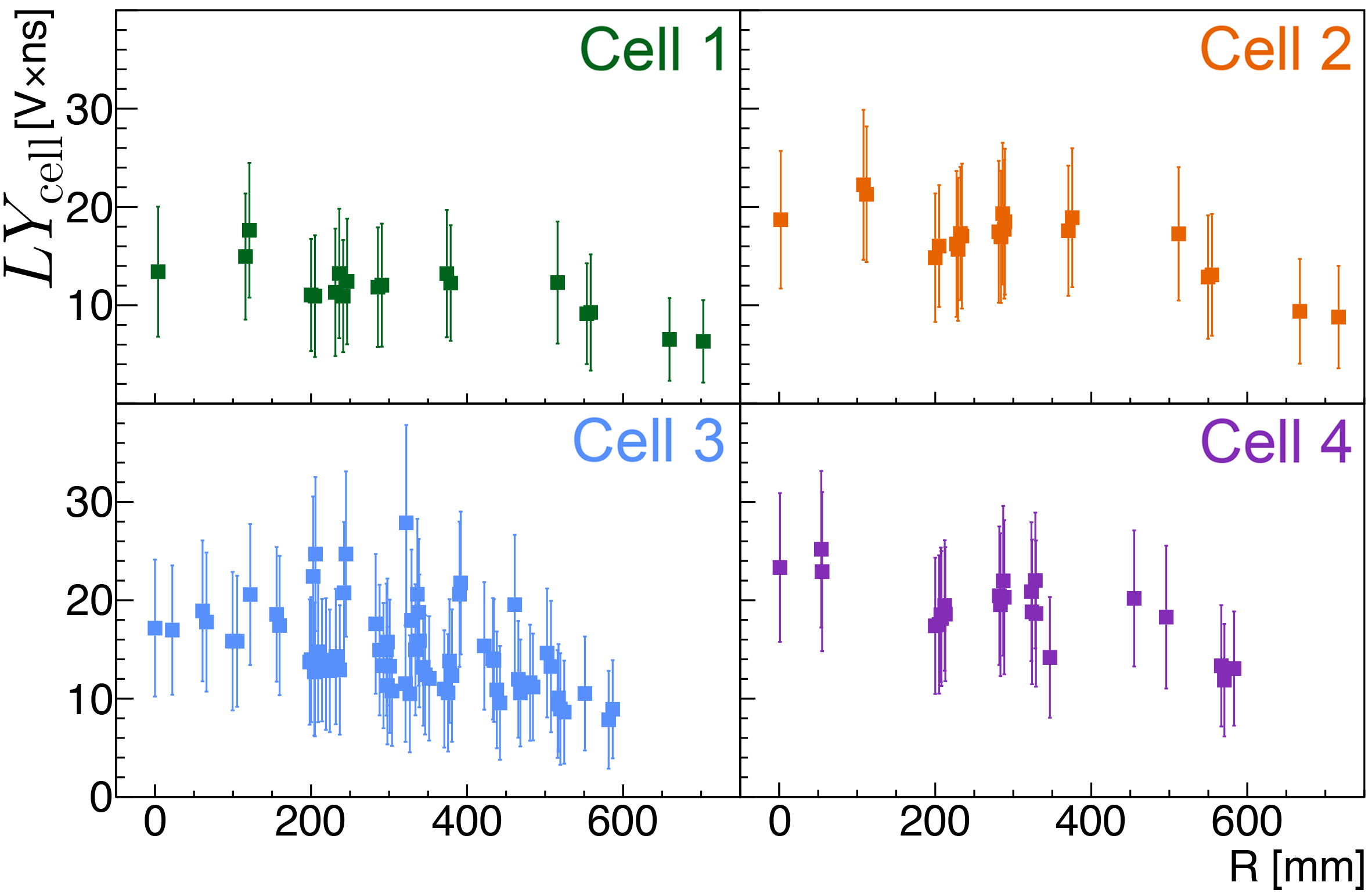}
    \caption{Mean value $LY_\text{cell}$ of the integrated light yield for each detector cell as a function of the distance~$R$ from the respective cell's centre. Only particle tracks at a beam incident angle of $\theta_X=0\degree$ and $\theta_Y=0\degree$ are taken into account. Cell~3 was studied at a larger number of different beam positions than the other cells (see Figure~\ref{fig:deglocations}). The error bars show the standard deviation of the distributions. Since $R$ does not provide an unambiguous definition of the exact beam position on the detector, the measured integrated light yields may differ between very similar values of $R$.}
    \label{fig:charge_allcells}
\end{figure}

At various particle crossing points on the detector (indicated in Figure~\ref{fig:deglocations}), the mean value of the integrated light yield for each cell ($LY_\text{cell}$) was evaluated with fixed beam incident angles $\theta_X=0\degree$ and $\theta_Y=0\degree$. The~measured mean values of $LY_\text{cell}$ and standard deviations of the $LY_\text{cell}$ distributions during the exposure to the 5\,GeV muon beam are shown in Figure~\ref{fig:charge_allcells} as a function of the radial distance $R$ between the particle crossing point and the centre of the respective detector cell. The mean values of $LY_\text{cell}$ are observed to strongly depend on the location where the muon crosses the detector cell: Most light is detected when the particle crossing point is close to a WOM, and all cells exhibit a decrease in integrated light yield for muons traversing the cells near their corners.

In smaller detector cells, the WOMs will on average be closer to the particle crossing point and thus to the location where the scintillation light is generated. This results in a higher expected light yield for smaller cells: The WOM covers a larger solid angle with respect to the particle track, and the photons need to travel a shorter distance, with consequently decreased probability of being absorbed in the LS or the cell walls before reaching the WOM. The LS cells of the prototype detector are of different size. The measured light yield is thus expected to vary depending on the cell: Based on their similar dimensions (see Figure~\ref{fig:cella}), we expected similar signals to be detected in the larger two cells (Cells~1 and~2), and in the smaller two cells (Cells~3 and~4), respectively, and that the light yield registered in the smaller detector cells (Cells~3 and~4) would be higher than that measured in the larger cells (Cells~1 and~2). As can be seen in Figure~\ref{fig:charge_allcells}, this was, however, not observed: The integrated light yields $LY_\text{cell}$ measured in Cells~1 and~3 (\textit{top~left} and \textit{bottom~left}) are generally lower than those detected in Cells~2 and~4 (\textit{top~right} and \textit{bottom~right}). The reason for this deviation was tracked down to imperfect optical coupling between one WOM and its SiPM array in each of Cells~1 and~3, resulting in significant signal losses for that WOM. Comparing the signals detected in Cells~1 and~3, and the signals in Cells~2 and~4, the measured integrated light yield $LY_\text{cell}$ is still higher for the respective smaller cells ($LY_\text{cell3} > LY_\text{cell1}$ and $LY_\text{cell4} > LY_\text{cell2}$). So in regards to size, the detectors perform as expected when taking into account the above-mentioned poor optical coupling for WOMs in Cells~1 and~3.

\FloatBarrier
\subsubsection{Likelihood-based correction of the detector response}
\label{Sec:DetectorResponseUniformityLikelihoodCorrection}

As described in Section~\ref{Sec:LightCollectionandDetectorResponseUniformity}, the detector response of a SBT cell will depend on the incident particle location, and a WOM close to the particle crossing point may measure an integrated light yield several times larger than that observed in the further WOM. In~\cite{Alt:2023vuu}, a likelihood-based procedure was developed to correct for this non-uniformity of the detector response which allows to reconstruct the particle crossing point on the detector cell, as well as its deposited energy. The same correction function is now applied to the data collected with the four-cells detector prototype at a beam incident angle of $\theta_X=0\degree$ and $\theta_Y=0\degree$ for all particle crossing locations shown in Figure~\ref{fig:deglocations}, using the centre of each cell at $R = 0$ mm as a reference point. 

The fractional integrated light yields in the \textit{upper} (or likewise \textit{lower}) WOM 
\begin{equation}\label{eq:LYF_WOM}
   f_{\text{WOM}_{u}} = \frac{LY_{\text{WOM}_{u}}}{LY_{\text{WOM}_u}+LY_{\text{WOM}_d}}
\end{equation}
and in each of their eight channels of five SiPMs
\begin{equation}\label{eq:LYF_ch}
    f_\text{ch} = \frac{LY_\text{ch}}{LY_\text{WOM}} \quad \textnormal{with} \quad \textrm{ch}= 1 ... 8
\end{equation} 
contain information on the $X$ and $Y$ coordinates of the particle crossing point on the detector. They are thus used as input variables to the likelihood-based correction, as is the difference in photon arrival times between the WOMs of each cell (see~\cite{Alt:2023vuu} for further details).

\begin{figure}[ht]
    \centering
    \includegraphics[width=0.85\textwidth]{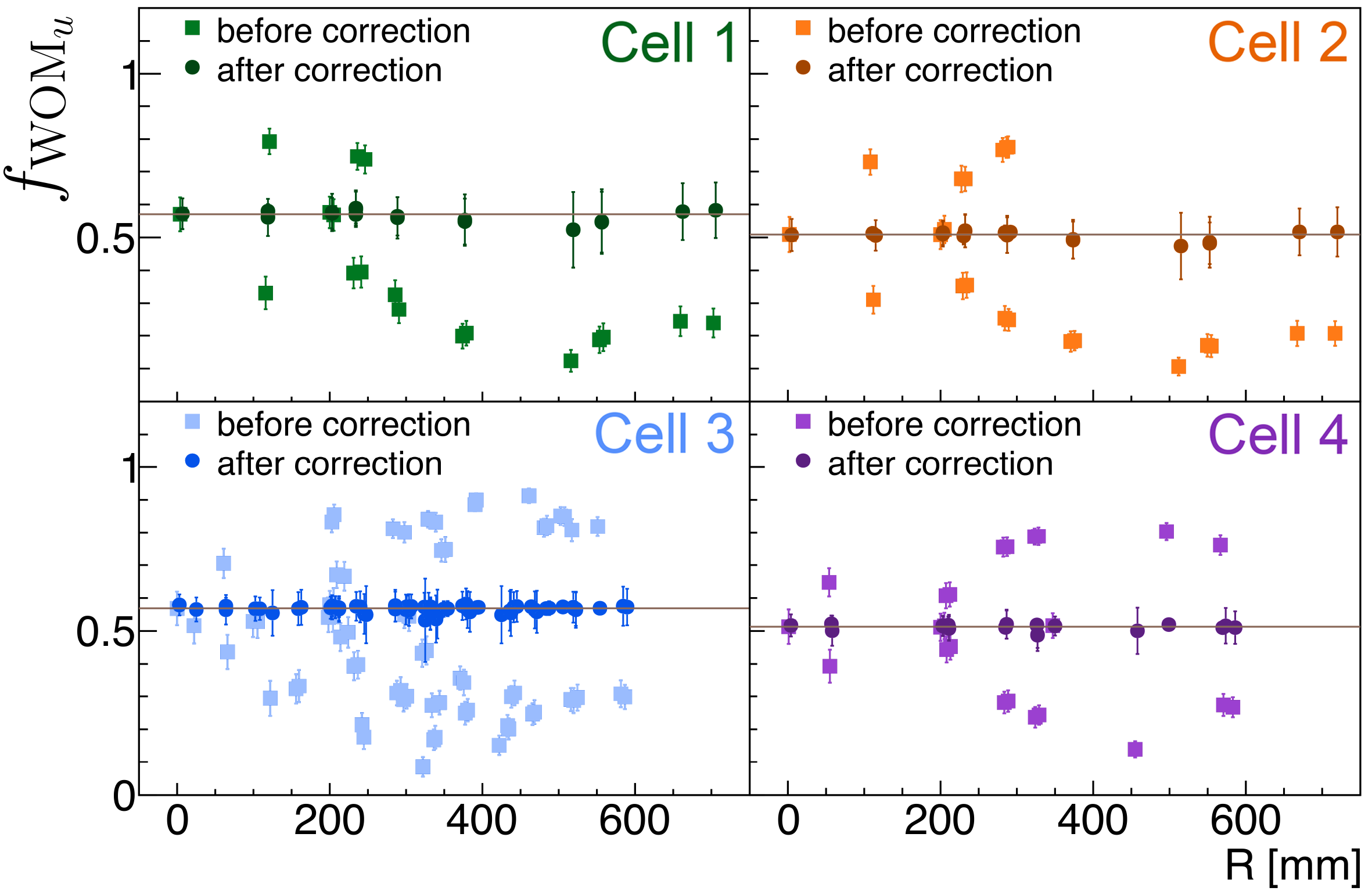}
    \caption{Fractional integrated light yield $f_{\text{WOM}_u}$ in the upper WOM for each detector cell as a function of distance $R$ from the respective cell's centre, both before and after applying the likelihood-based correction. Only particle tracks at a beam incident angle of $\theta_X=0\degree$ and $\theta_Y=0\degree$ are taken into account. The error bars show the standard deviations of the distributions. The \textit{brown} line represents the value measured at the reference point at $R = 0$\,mm.    
    \label{fig:likelihood_wom1frac}}
\end{figure}

For $f_{\text{WOM}_{u}}$, the results of this correction for each detector cell are shown in Figure~\ref{fig:likelihood_wom1frac}. The~correction significantly reduces the deviation of the values of $f_{\text{WOM}_u}$ from the reference value measured at $R = 0$\,mm. The performance of this likelihood-based correction is observed to be similar for all detector cells, with no strong impact from cell size or magnitude of the integrated light-yield fraction. While the deviation of the integrated light yield fraction remaining after the corrections and its increased standard deviation indicates that not all particle crossing points in all events could be correctly reconstructed, this result proves the success of the correction in equalising the integrated light yield of both WOMs in a detector cell.

\FloatBarrier
\subsubsection{Comparison to simulation}
\label{Sec:DetectorResponseUniformityDataSimulationComparison}

This section provides a qualitative comparison of measurements and simulation, for a more detailed description of the detector Monte Carlo simulation see~\cite{Alt:2023vuu} and~\cite{lyons2025}. The exact geometry of the four-cell detector prototype was implemented in GEANT4~\cite{Agostinelli} and used simulate data with a 5\,GeV muon beam at an incident angle of $\theta_X=0\degree$ and $\theta_Y=0\degree$ and the same locations on detector Cell~4 as those measured during the test beam. 

Several sets of data were simulated for uniform inner cell surface reflectivities between 50\% and 100\% of the Optopolymer\textsuperscript{\textregistered} BaSO$_4$ paint's nominal wavelength-dependent reflectivity, which is specified to be better than 95\% in the UV and visible range according to the manufacturer~\cite{baso4}. These were then compared to the measurements~\cite{lyons2025} of this test beam exposure. 

In order to identify the value of reflectivity bringing simulation and measurements in closest agreement, we defined a~$\chi^2$~function to compare the average photon yields in the simulation ($s_i$) to the measured average integrated light yield ($d_i$) for all particle crossing points ($i = 1, ..., N$). A~variable scale parameter~$\lambda(\alpha)$, depending on the relative reflectivity $\alpha$, was applied to the simulated photon yields to enable a direct comparison of simulated and measured data. For any simulated reflectivity in steps of 5\%, the $\chi^2$ between these distributions was then summed over all particle crossing points $i$ and minimised:

\begin{equation}\chi_\textrm{Tot}^2\left(\lambda(\alpha), \alpha\right) = \sum_i\frac{(\lambda(\alpha)\cdot s_i(\alpha) - d_i)^2}{\sigma_i^2},
\end{equation}

with $\sigma_i$ being the standard deviation of the simulated photon yield distribution divided by the square root of number of events.

\begin{figure}[ht]
    \centering
    \includegraphics[width=0.7\textwidth]{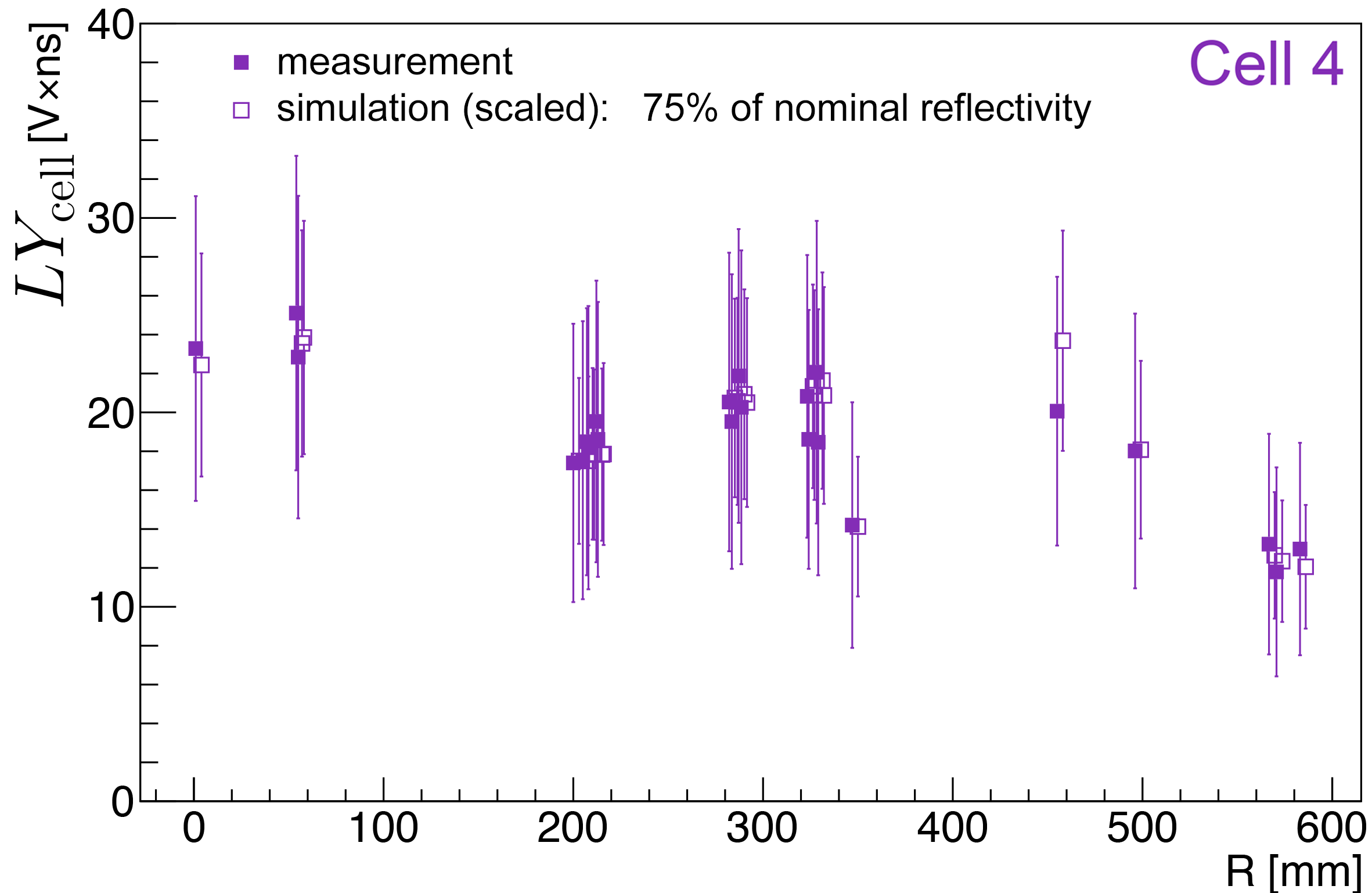}
    \caption{Average integrated light yield $LY_\text{cell}$ measured in Cell~4 for various particle crossing points on the detector at a beam incident angle of $\theta_X=0\degree$ and $\theta_Y=0\degree$ as a function of distance $R$ from the cell centre, compared to the results of the GEANT4 simulation with a muon beam energy of 5\,GeV. Error bars show the standard deviations of the distributions. The cell wall reflectivity applied in the simulation to best reproduce the measured data is 75\% of the nominal values provided by the manufacturer of the BaSO$_4$ paint.}
    \label{fig:charge_cell4_sim}
\end{figure}

The value of $\lambda$ yielding minimal $\chi^2$ was then chosen as the best scale parameter for a given value of $\alpha$, and the lowest $\chi^2$ values for each $\alpha$ were compared: The overall minimum in $\chi^2_\textrm{Tot}$ was reached for a relative reflectivity $\alpha = 75$\% of the nominal manufacturer specification~\cite{lyons2025}, corresponding to a total reflectivity of 73\% at 380\,nm. Figure~\ref{fig:charge_cell4_sim} shows the comparison of the measured average integrated light yields in Cell~4 and the average photon yields simulated for a relative cell wall reflectivity of $\alpha=75\%$ of the nominal spectrum, exhibiting good agreement for all particle crossing points on the detector cell.

It is concluded that, while the simulation was very successful in reproducing the detector response at various particle crossing points, the nominal reflectivity of the BaSO${_4}$ paint had not been reached. This is likely due to the increased difficulty of coating the inner walls of the prototype cells after welding via the small WOM openings and by reactions between the paint and the COR-TEN\textsuperscript{\textregistered} steel~\cite{deucher2026}.

\FloatBarrier
\subsubsection{Detector response for particles crossing multiple cells}
\label{Sec:DetectorResponseParticleaCrossingSeveralCells}
\begin{figure}[ht]
    \centering
    \includegraphics[width=0.6\textwidth]{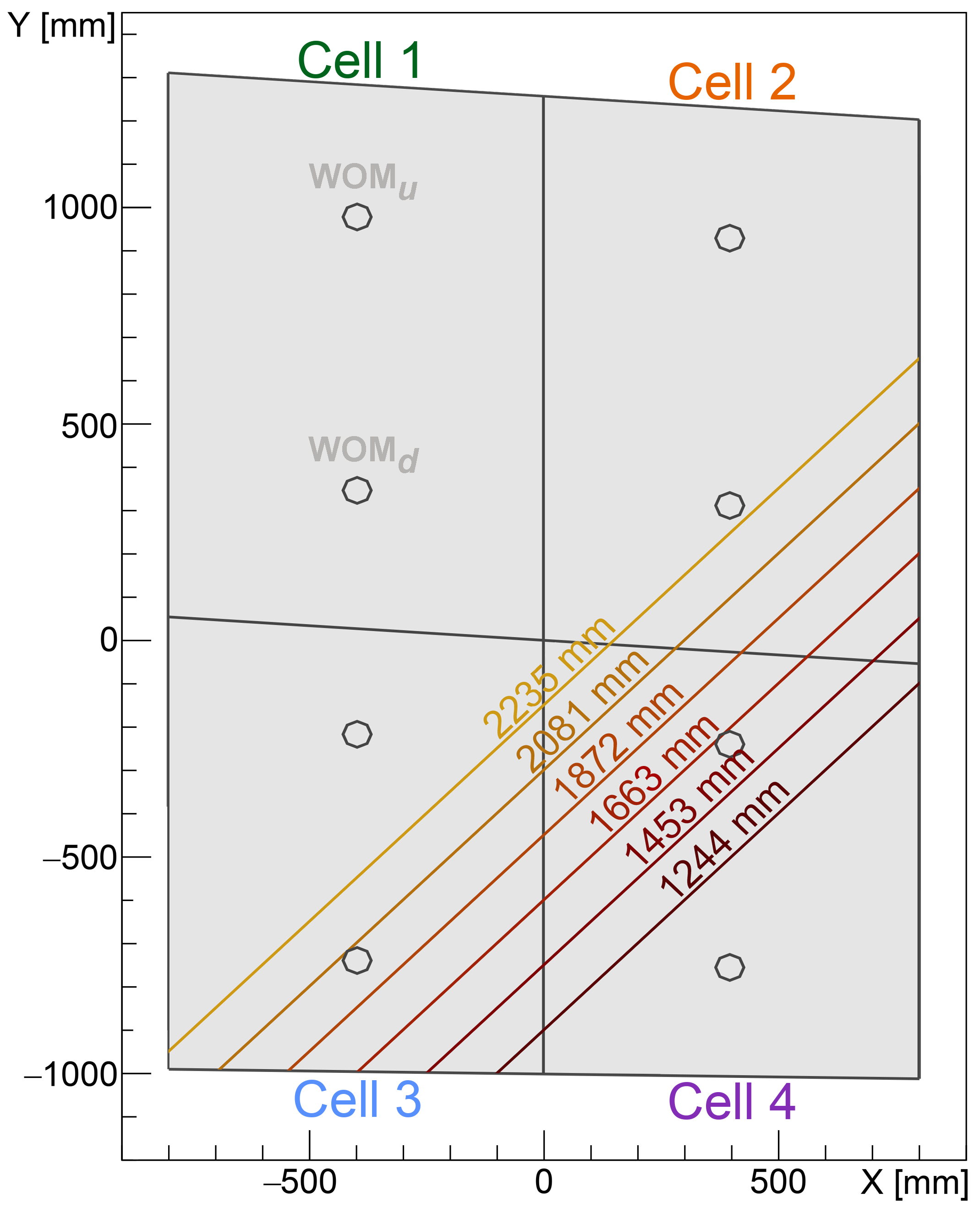}
    \caption{Trajectories of particles crossing the central plane ($Z=0$) of the four-cell detector at a beam incident angle of $\theta_X=90\degree$ and $\theta_Y=45\degree$. Detector cell walls and WOM locations are shown by outlines. The numbers indicate the respective track length of the particles inside the liquid scintillator, depending on the beam's incident location.}
    \label{fig:tracks_rotx90roty45}
\end{figure}

In addition to measurements conducted with the beam perpendicularly crossing a single detector cell at shortest path length ($\theta_X=0\degree$ and $\theta_Y=0\degree$), also particles traversing multiple detector cells were recorded. Figure~\ref{fig:tracks_rotx90roty45} shows such tracks through the detector at an incident angle of $\theta_X=90\degree$ and $\theta_Y=45\degree$ with respect to the normal to the detector front plate, providing a more realistic emulation of the particle trajectories expected in the SHiP experiment. These particles all cross Cells~2,~3,~and~4 at the central plane of the detector, but have different track lengths in each cell, as well as varying total track length, depending on their location of incidence. 

The integrated light yield measured in a detector cell $LY_\text{cell}$ is expected to be dependent on the length of the particle track within the liquid scintillator, as a longer track length through the LS will result in higher energy deposition, and consequently the emission of more scintillation light. As~demonstrated for beam incident angles of $\theta_X = 0\degree$ and $\theta_Y = 0\degree$ in Section~\ref{Sec:LightCollectionandDetectorResponseUniformity}, the measured integrated light yield will also depend on the distance between the particle's crossing point and the WOMs, and consequently the location at which the particle crosses a cell at a given angle.

\begin{figure}[htb]
    \centering
    \includegraphics[width=0.7\textwidth]{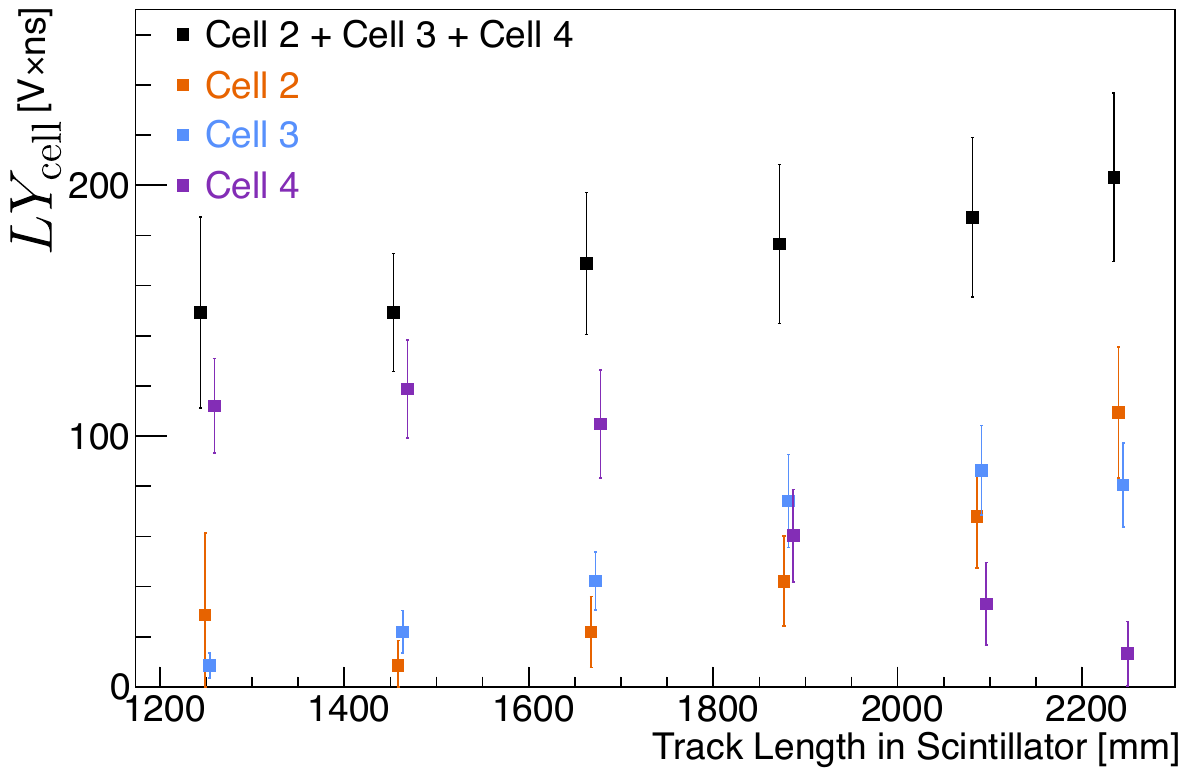}
    \caption{Integrated light yield $LY_\text{cell}$ summed over both WOMs of each detector cell for the particle trajectories shown in Figure~\ref{fig:tracks_rotx90roty45}. The sum of the integrated light yields measured in all traversed cells (Cells~2,~3, and~4) is also shown. An upper threshold of 100\,V\,$\times$\,ns per WOM is applied to reject events with a very high number of detected photons (e.g. from hadronic showers in the detector caused by hadrons from beam contamination) where the amplified SiPM signal is saturated. Error bars indicate the standard deviations of the distributions.}
    \label{fig:charge_callcells_rotx90roty45}
\end{figure}

Figure~\ref{fig:charge_callcells_rotx90roty45} shows the integrated light yield of both WOMs of each detector cell $LY_\text{cell}$ as a function of total track length within the LS, as well as the summed signal over Cells~2,~3, and~4 for particles on the trajectories indicated in Figure~\ref{fig:tracks_rotx90roty45}. As expected, a longer particle track length inside a detector cell resulted in a larger integrated light yield measured for this cell. 
In these measurements, the trajectories with the shortest total track lengths in LS mostly crossed Cell~4, resulting in large measured integrated light yields in this cell and smaller ones in Cells~2 and~3. With increasing total track length, also the integrated light yields measured in Cells~2 and~3 rose, while the integrated light yield detected in Cell~4 decreased. It is furthermore observed that, when the particle crossed a detector cell in close vicinity to a WOM (i.e., for Cell~2 at the total track length of 2235\,mm and for Cell~3 at the total track length of 2081\,mm), the detected integrated light yield was highest and could even surpass the one of longer particle tracks through the same cell (compare e.g. the integrated light yields detected in Cell~3 for total track lengths of 2235\,mm and 2081\,mm).

\FloatBarrier
\subsection{Time response}
\label{Sec:TimeResponse}

To study the time response of the detector, the signal arrival times at both WOMs of a cell were estimated via Constant Fraction Discrimination~(CFD): For each event, the waveforms measured in the eight SiPM channels of each WOM were summed to minimise fluctuations caused by electronic noise. To the resulting sum of waveform a smoothing was applied, using a Gaussian kernel extending over $\pm$3$\sigma$, with $\sigma=2$\,ns and a bin width of 0.321\,ns. In each bin, the smoothed value was calculated as a weighted average of neighbouring bins using Gaussian weights, with the normalisation adjusted near the array boundaries. Finally, the time at which this sum signal reached 25\% of its maximum value was then determined via linear interpolation.

For each cell, the resulting signal arrival times ${T}_{u}$ and ${T}_{d}$ measured at WOM$_{u}$ and WOM$_{d}$, respectively, were corrected by the time $T_\textrm{ref}$ given by the small scintillator at the centre of the upstream Beam Telescope:  $T_{u}^\textrm{corr} = {T}_{u}-{T}_\textrm{ref}$ and ${T}_{d}^\textrm{corr} = {T}_{d}-{T}_\textrm{ref}$. Using the same WaveCatcher, the waveform of this scintillator was recorded simultaneously with the WOM signals, and again a CFD threshold of 25\% was applied. Uncertainties in the timing of the WaveCatcher clock were thus cancelled. 

\begin{figure}[hbt]
    \centering
    \begin{subfigure}[c]{0.55\textwidth}
        \includegraphics[height=7.0cm]{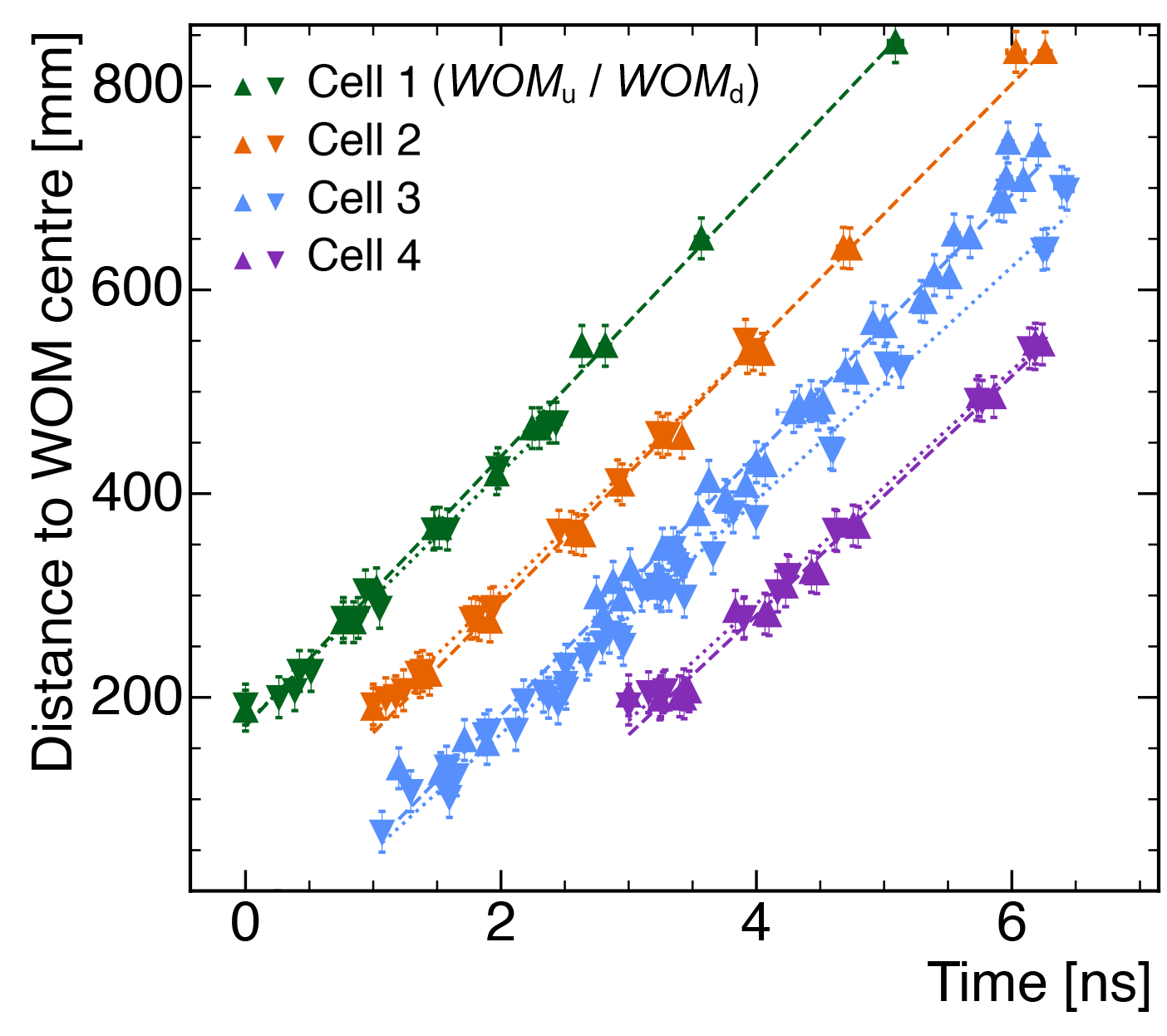}
        \caption{Distance to WOM vs. signal arrival time.}
        \label{fig:allcells_time_vs_distance}
    \end{subfigure}   
    \hspace{1.0mm}
    \begin{subfigure}[c]{0.42\textwidth}  
        \includegraphics[height=6.2cm]{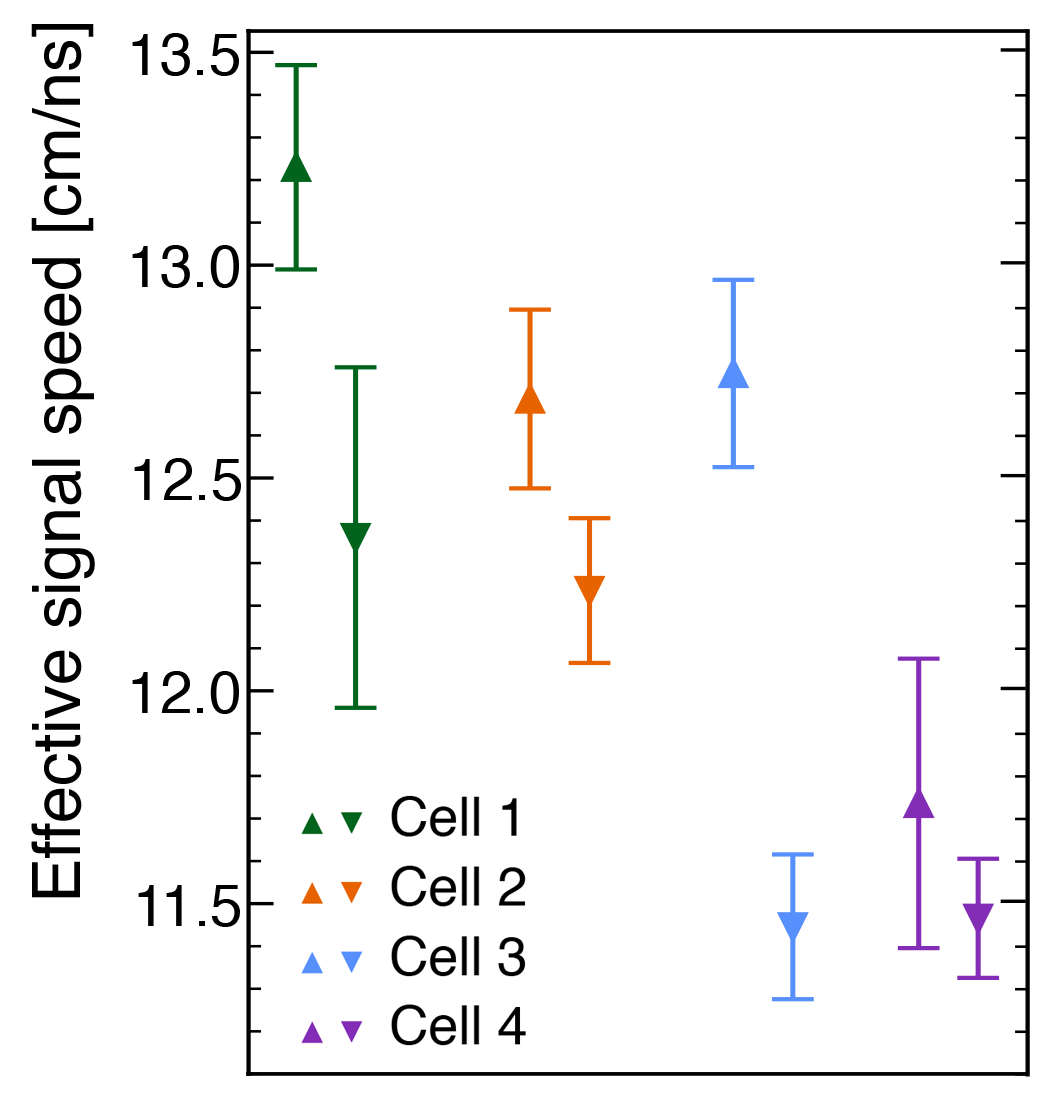}
        \vspace{7mm}
        \caption{Effective signal speed.}
        \label{fig:effective_signal_speed}
        \end{subfigure}    
    \caption{(a) Distance between particle crossing location for $\theta_X=0\degree$ and $\theta_Y=0\degree$ in all four detector cells (see Figure~\ref{fig:deglocations}) and the respective WOM centre as a function of mean of the corrected signal arrival time distributions $\bar{T_{u}}^\textrm{corr}$~or~$\bar{T_{d}}^\textrm{corr}$ (indicated by the directions of the triangles). For better readability, the plots for each detector cell are shifted by 1\,ns relative to each other. Positions where one WOM blocks the direct path for photons from the respective interaction to the other WOM have been excluded. (b) Slopes of the linear fits to (a): This quantity is interpreted as the effective signal speed in a given detector cell.}
    \label{fig:time_signal_speed}
\end{figure}

Depending on the location of interaction between incoming particle and detector, the generated scintillation light will take different paths to the WOMs. Detector simulations for a standard SBT cell estimate an average of twelve reflections on its inner walls (with resulting path lengths of up to 180\,cm) before a scintillation photon is either converted on the WOM surface into a secondary wavelength-shifted photon or absorbed by other detector material~\cite{Alt:2023vuu,Brignoli:2025spc}. For~a beam incident angle of $\theta_X=0\degree$ and $\theta_Y=0\degree$ and using the measurements on all four detector cells shown in Figure~\ref{fig:deglocations}, the relation of the distance between particle crossing location and centre of the respective WOM and the mean of the corrected signal arrival time distributions ($\bar{T_{u}}^\textrm{corr}$~or~$\bar{T_{d}}^\textrm{corr}$) is shown in Figure~\ref{fig:allcells_time_vs_distance}. The slopes of the linear fits to these plots represent the effective signal speed in the respective detector cell, see Figure~\ref{fig:effective_signal_speed}. At about 12.5\,cm/ns, it is similar for all four detector cells, with some differences likely caused by the variation in cell geometry.

\begin{figure}[ht]
	\centering
	\includegraphics[width=0.85\textwidth]{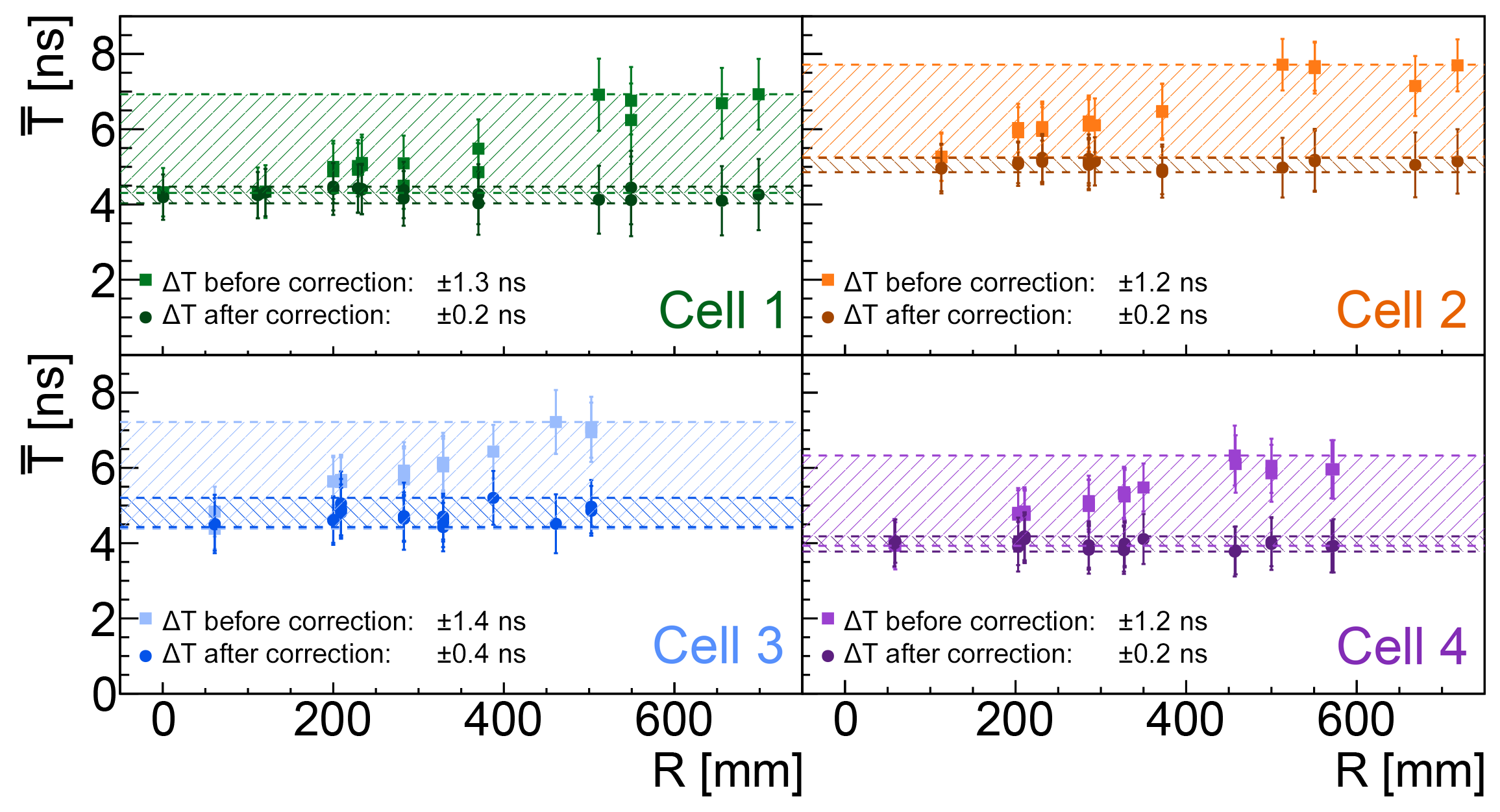}
	\caption{Mean $\bar{T}$ of the average signal arrival times, shown for a beam incident angle of $\theta_X=0\degree$ and $\theta_Y=0\degree$, as a function of the distance~$R$ between the particle crossing point and the centre of the respective detector cell, both before and after applying the likelihood-based correction described in Section~\ref{Sec:TimeResponseLikelihoodCorrection}. The uncertainties are given by the standard deviations of the $\bar{T}$ distributions. Cross-hatched bands indicate the time variation across the entire detector cell before (\textit{right-diagonal}) and after the likelihood correction~\hbox{(\textit{left-diagonal})}.}	
	\label{fig:average_time_00_corr_chToT}
\end{figure}

The average signal arrival time was calculated for each event as \hbox{$T_{ud}=\frac{1}{2}(T_{u}^\text{corr}+T_{d}^\text{corr})$}. The~mean values $\bar{T} = {\bar{T}}_{ud}$ of the $T_{ud}$ distributions for each detector cell are shown in Figure~\ref{fig:average_time_00_corr_chToT} as a function of the distance $R$ between the particle crossing point and the centre of the respective cell. Before applying further corrections, they are observed to vary within $\pm$1.2\,ns (Cell~4) to $\pm$1.4\,ns~(Cell~3). 
 
For individual locations on the detector cells, the intrinsic time resolutions are given by the standard deviations of the $T_{ud}$ distributions and shown in Figure~\ref{fig:average_time_00_corr_chToT} as error bars. These intrinsic time resolutions vary between 0.6\,ns and 1.0\,ns, depending on the distance $R$ between the particle crossing point and the centre of the respective cell. Since the time resolution of $T_\text{ref}$ is about 0.2\,ns and thus significantly better than the time resolution of $T_{ud}$, no further smearing effect from $T_\text{ref}$ needs to be applied. 

\FloatBarrier
\subsubsection{Likelihood-based correction of the time response}
\label{Sec:TimeResponseLikelihoodCorrection}

Analogous to the method used in Section~\ref{Sec:DetectorResponseUniformityLikelihoodCorrection}, also the variation in detector response time can be reduced by applying a likelihood-based correction. Based on a set of variables extracted from the measured waveforms, all data taken at a beam incidence angle of $\theta_X=0\degree$ and $\theta_Y=0\degree$ were corrected with respect to the reference position at the centre of the respective detector cell.

A first approach applied the fractional integrated light yield $f_{\text{WOM}_u}$ in the \textit{upper} WOM \hbox{(Equation~\eqref{eq:LYF_WOM})} and $f_\text{ch}$ in each group of SiPMs (Equation~\eqref{eq:LYF_ch}), as well as the difference in signal arrival times between both WOMs of a cell. This procedure allowed to reduce the uncertainty on 
the average signal arrival times from~$\pm$1.2\,ns to~$\pm$0.2\,ns (Cell~4) and from~$\pm$1.4\,ns to~$\pm$0.3\,ns~(Cell~3). 

In a second method, the likelihood-based correction used the Time-over-Threshold~(ToT) for each channel (in this case with a general fixed threshold of 10\,mV) instead of its fractional integrated light yield~$f_\text{ch}$. With a performance comparable to the correction taking into account the integrated light yields, the uncertainty on the average signal arrival times was reduced to ranges of~$\pm$0.2\,ns~(Cell~4) and~$\pm$0.4\,ns (Cell~3), see Figure~\ref{fig:average_time_00_corr_chToT}. 

\begin{figure}[ht]
    \centering
    \includegraphics[width=0.7\textwidth]{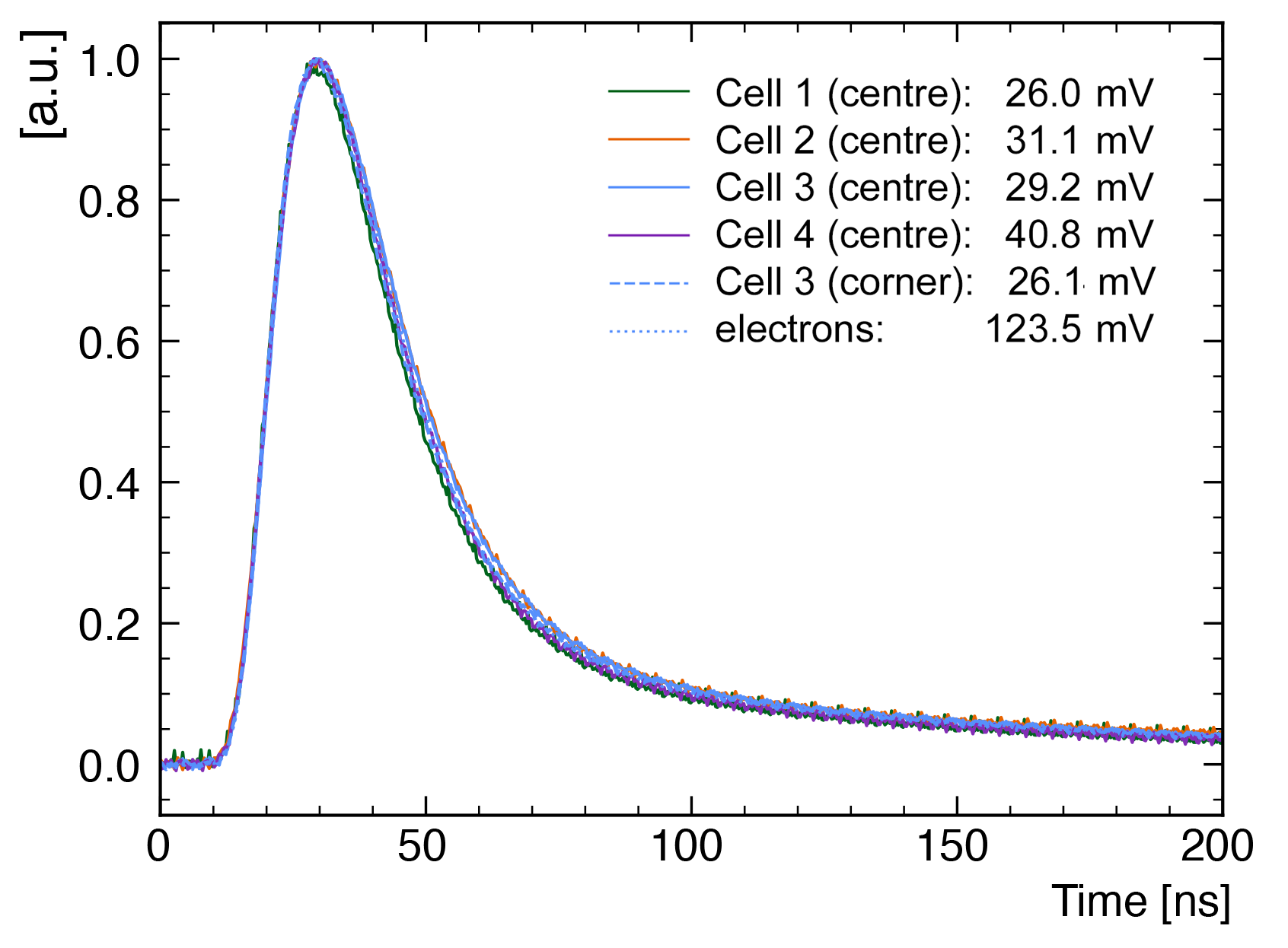}
    \caption{Average of 5\,000 waveforms collected in channel~0 of WOM$_{u}$ of detector Cells~1 -- 4 for a beam incident angle of $\theta_X=0\degree$ and $\theta_Y=0\degree$, a muon beam position on the \textit{centre} or \textit{corner} (Cell~3) of the cell, and an electron beam positioned at the \textit{centre} of Cell~3. The waveforms have been normalised to the same integrated light yield and aligned at the rising edge at 50\,\% of their measured maximum amplitude given in the legend.}
    \label{fig:average_waveform}
\end{figure}

This is a highly important result: In the final detector configuration of the SHiP SBT, the full waveforms of each channel are unlikely to be sampled, while instead only timestamp and ToT are expected to be read out in order to massively reduce data volume and acquisition time. This readout strategy is further encouraged by a comparison of the average waveforms measured in channel~0 of WOM$_{u}$ for different detector cells \hbox{(Cells~1 -- 4)}, different points of particle interaction within the cell \hbox{(centre vs.\ corner)}, or even different particle types \hbox{(muons vs.\ electrons)}, see Figure~\ref{fig:average_waveform}: Normalised to the same integrated light yield and aligned by the rising edge, the shape of all waveforms is in close agreement and can be well characterised by a measurement of its maximum amplitude and Time-over-Threshold. 

\FloatBarrier
\subsubsection{Time response for inclined tracks}
\label{Sec:TimeResponseDataSimulation}

\begin{figure}[htb]
	\centering
	\includegraphics[width=0.6\textwidth]{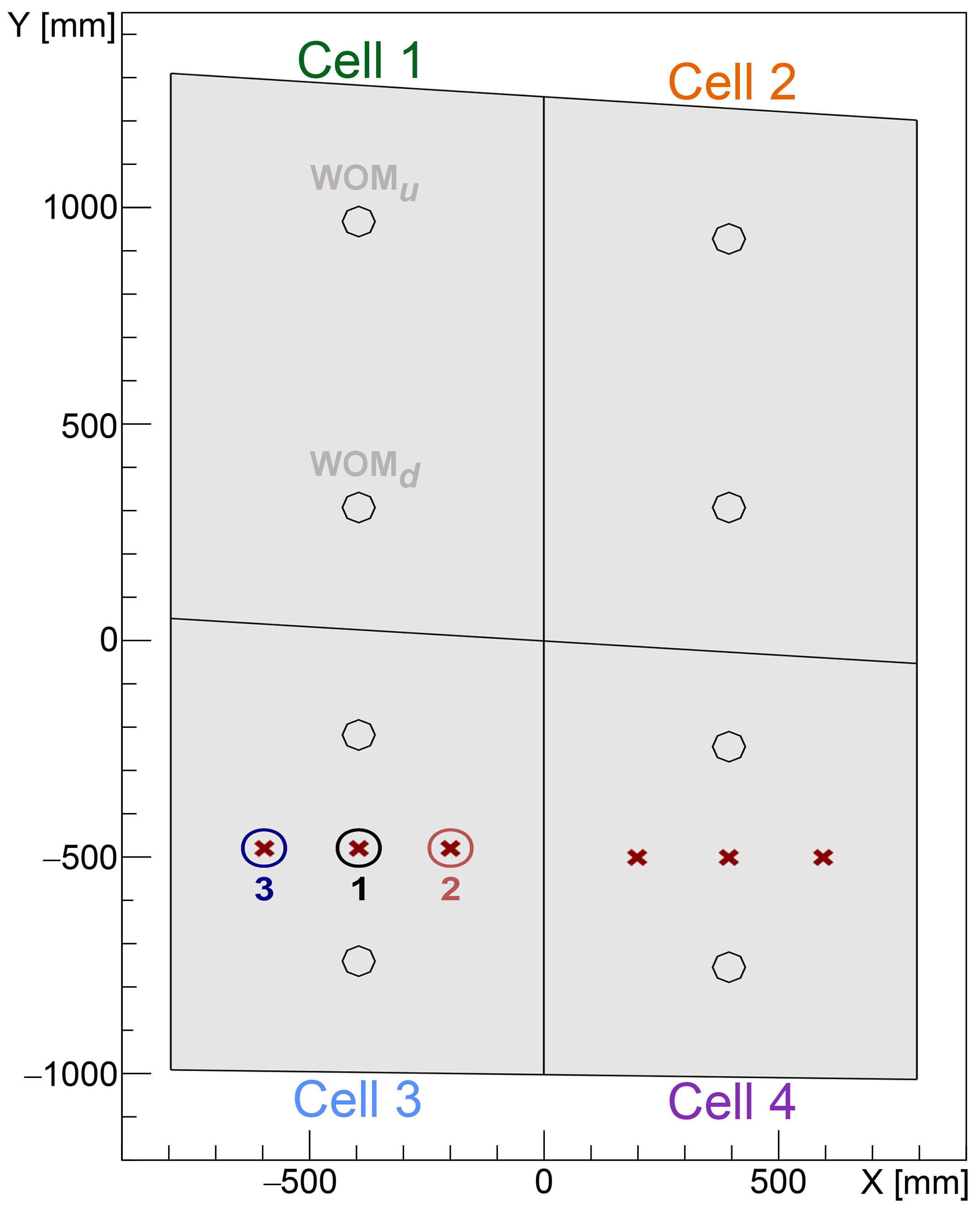}
	\caption{Locations of particle crossing points on the central plane of the detector, set between the WOMs of detector Cells~3~and~4, that were used in the time response analysis for differently inclined tracks: For~each of these locations, the vertical axis of the detector was fixed at $\theta_Y=0\degree$, while the detector was rotated in steps of 15$\degree$ around the horizontal axis, between $\theta_X=15\degree$ and $\theta_X=90\degree$ . The circled points in Cell~3 were furthermore studied in the likelihood-based correction described in Section~\ref{Sec:ReconstructionAngles}.}	
	\label{fig:Grid_3pos}
\end{figure}

A change in inclination of a particle trajectory on the detector cell will result in a change of the track length in the liquid scintillator and of the locations of scintillation light emission. This influences not only the integrated light yield, but also the signal arrival time. To study and quantify this effect, measurements were taken at six different beam locations on the detector (three different position in $X$ for each of Cells~3 and~4, at $Y \approx -500$\,mm for equal distances to WOM$_u$ and WOM$_d$) for various incident angles ($\theta_X$ was varied in steps of $15\degree$ from $\theta_X=15\degree$ to $\theta_X=90\degree$, at fixed $\theta_Y=0\degree$), see Figure~\ref{fig:Grid_3pos}. The vertical position of the detector was adjusted to keep $Y$ constant despite the variations in $\theta_X$.

\begin{figure}[htb]
    \centering
    \begin{subfigure}[c]{0.46\textwidth}
        \includegraphics[height=6.6cm]{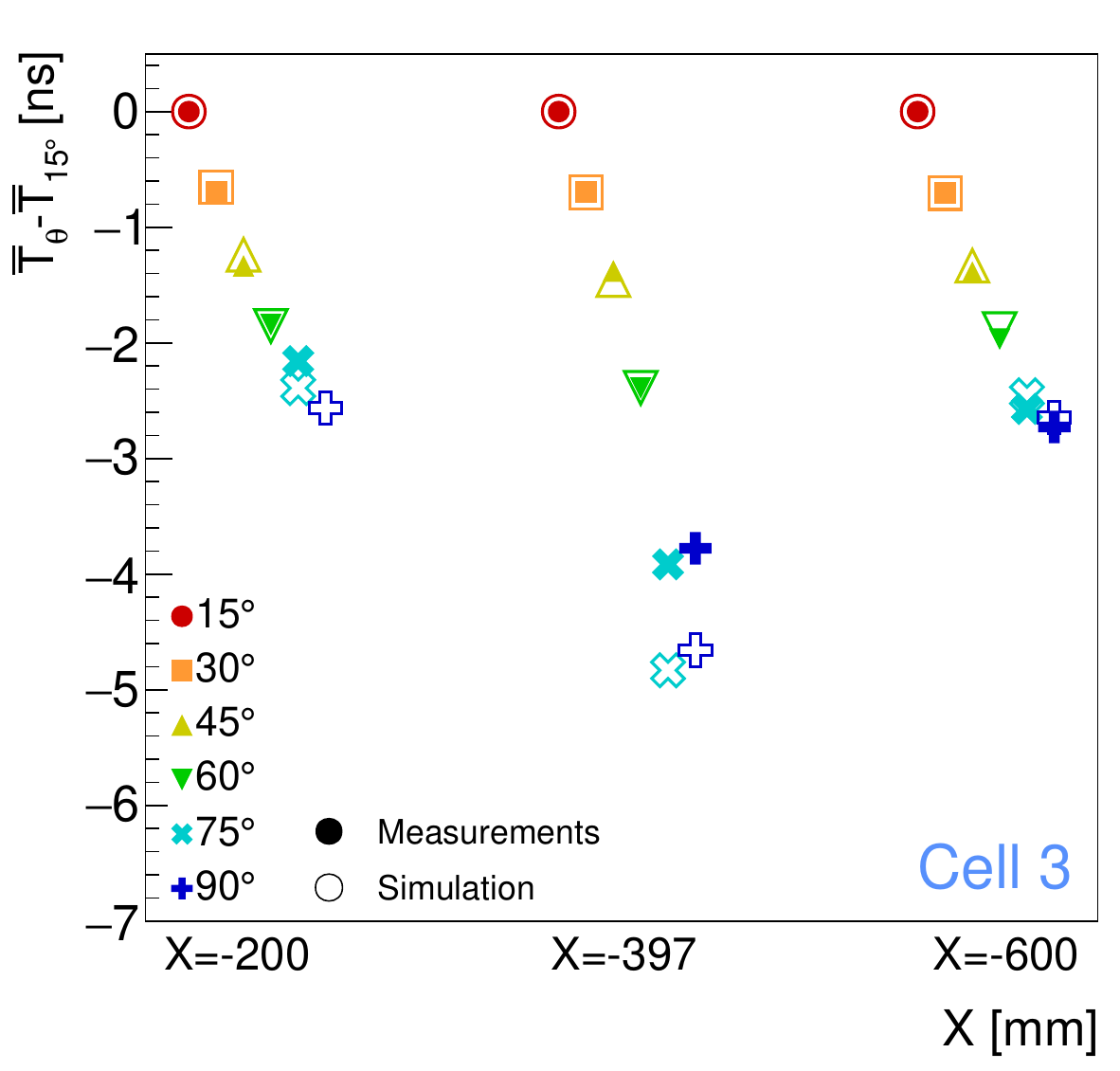}
        \caption{Difference of signal arrival times in Cell~3.}
        \label{fig:time_variation_angles_a}
    \end{subfigure}   
    \hspace{5.0mm}
    \begin{subfigure}[c]{0.46\textwidth}
        \includegraphics[height=6.6cm]{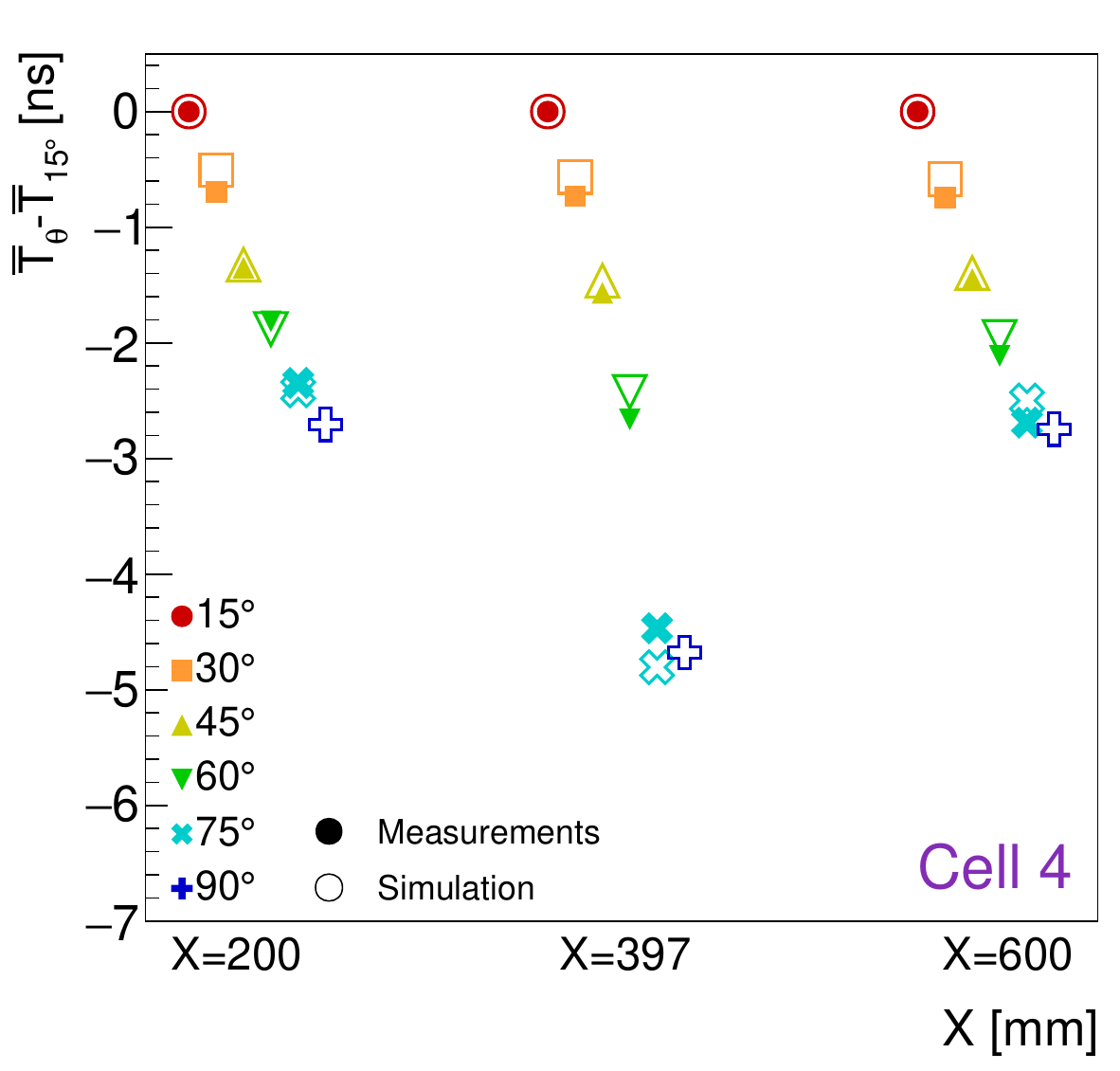} 
        \caption{Difference of signal arrival times in Cell~4.}
        \label{fig:time_variation_angles_b}
    \end{subfigure}    
    \caption{Difference of the mean values $\bar{T}$ of the average signal arrival times between tracks with incident angles of $\theta_X=15\degree$ and other inclinations between $\theta_X=30\degree$ and $\theta_X=90\degree$. $\bar{T}$ was measured in detector Cells~3~(a) and~4~(b) for particles crossing the respective detector cell at three different locations between the WOMs, 
    at $Y \approx -500$\,mm for equal distances to WOM$_u$ and WOM$_d$, see Figure~\ref{fig:Grid_3pos}. While $\theta_X$ was varied between $\theta_X=15\degree$ and $\theta_X=90\degree$ in steps of $15\degree$, rotation around the vertical axis was fixed at $\theta_Y=0\degree$. The uncertainties are obtained by error propagation of the uncertainties on $\bar{T}_{\theta}$ and $\bar{T}_{15\degree}$ from a Gaussian fit on the respective distributions. They are typically of the order of 20\,ps, smaller than the plotted marker sizes and therefore not visible. For better readability, measured points at different $\theta_X$ are slightly shifted in $X$.}
    \label{fig:time_variation_angles}
\end{figure}

Figure~\ref{fig:time_variation_angles} shows a comparison of the mean values $\bar{T}$ of the average signal arrival times in measured and simulated data for Cell~3~(\ref{fig:time_variation_angles_a}) and Cell~4~(\ref{fig:time_variation_angles_b}): At each of the positions indicated in Figure~\ref{fig:Grid_3pos}, and for each angle of rotation in $\theta_X$, 
the difference $\bar{T}_{\theta} - \bar{T}_{15\degree}$ between the means of the signal arrival times at angle
$\theta$ and 15$\degree$ is calculated. ($\theta_X=15\degree$ was chosen as the reference, as no data had been recorded at position~3 with $\theta_X=0\degree$ and $\theta_Y=0\degree$.) The largest variations in signal arrival time (up~to~5\,ns) are observed in the centre of the detector.

The data analysis followed the same steps as described in Section~\ref{Sec:TimeResponseLikelihoodCorrection}. For the simulation, the signal arrival times were derived using the arrival times of photons at the SiPMs, relative to the start time of the muon. These simulated arrival times were then used to produce the digitised signal by stacking single-photon waveforms from dedicated SiPM dark-count measurements. After digitisation, the simulated waveforms were analysed using the same methods as were applied for the test beam measurements. 

In general, measurement data and simulation results are in very good agreement, with deviations of at most 0.2\,ns. Significant discrepancies are only observed at the central position (particularly in Cell 3 at $X = -397$\,mm) for larger incident angles of $\theta_X=75\degree$~and $\theta_X=90\degree$. These discrepancies are likely caused by a saturation of the eMUSIC signal which can occur when a beam particle directly traverses one or both WOMs, resulting in a large number of detected photons that is currently not modelled in the simulation.

\FloatBarrier
\subsection{Reconstruction of particle crossing coordinates and incident angles}
\label{Sec:Reconstruction}
In the SHiP experiment, the coordinates and incident angles of particles traversing the cells of the SBT are a-priori unknown. To~perform a more comprehensive study of the SBT's capability to reconstruct particle crossing coordinate and track inclinations, the likelihood-based reconstruction method presented in Section~\ref{Sec:DetectorResponseUniformityLikelihoodCorrection} was applied to a larger variety of beam positions and incident angles.

\FloatBarrier
\subsubsection{Incident angles and angular resolution}
\label{Sec:ReconstructionAngles}

For the three positions in detector Cell~3 (marked \textit{1}, \textit{2}, \textit{3} in Figure~\ref{fig:Grid_3pos}) at $Y \approx -500\,\text{mm}$, data was taken at different beam incident angles: The detector was rotated around the horizontal axis from $\theta_X = 0\degree$ to $\theta_X = 90\degree$, in steps of 15$\degree$ at fixed vertical angle $\theta_Y=0\degree$, and~around the vertical axis from $\theta_Y = 0\degree$ to $\theta_Y = 90\degree$, in steps of 15$\degree$ at fixed horizontal angle $\theta_X=0\degree$. For each position, the likelihood-based reconstruction was applied to its complete dataset containing tracks of all angles. This analysis was performed twice for the central Position~1: \textit{Analysis~1} used only information from Cell~3 and Cell~4, while \textit{Analysis~2} used the data of all four detector cells -- even if a particle does not cross a specific detector cell, the absence of a measured signal in this cell helps to constrain the possible angular range in the track reconstruction.

\begin{figure}[hbt]
    \centering
    \begin{subfigure}[c]{0.46\textwidth}        
        \includegraphics[height=6.6cm]{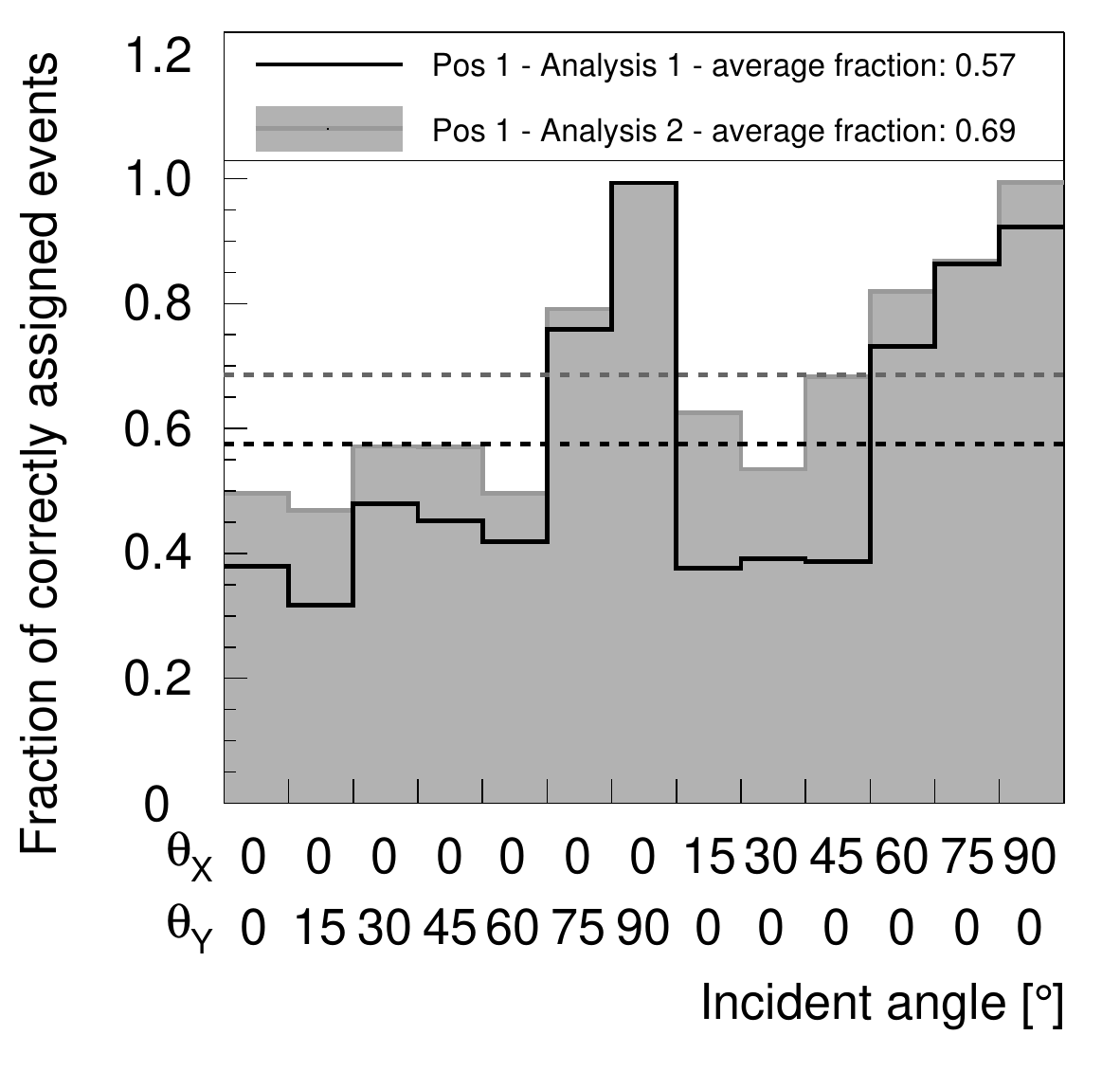}
        \caption{Analyses~1 and~2 for Position~1.}
        \label{fig:perc_corr_3pos_a}
    \end{subfigure}   
    \hspace{5.0mm}
    \begin{subfigure}[c]{0.46\textwidth}  
        \includegraphics[height=6.6cm]{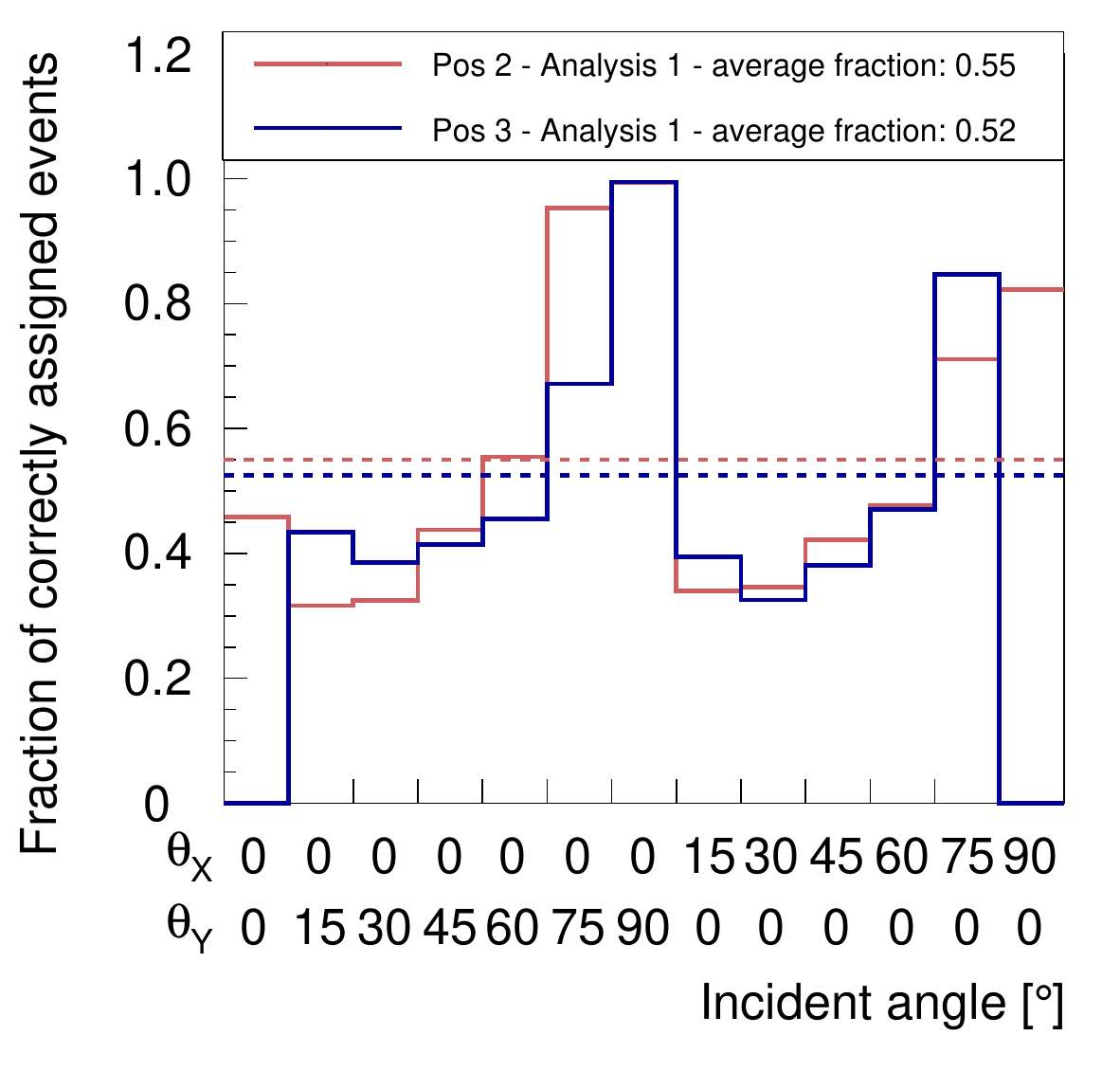}
        \caption{Analysis~1 for Positions~2 and~3.}
        \label{fig:perc_corr_3pos_b}
    \end{subfigure}    
    \caption{Fraction of events with correctly identified incident angles ($\theta_X$ and  $\theta_Y$) by the likelihood-based correction. Dashed lines represent the average over all inclinations. (a) Comparison of Analysis~1 (using only information from Cell~3 and Cell~4) and Analysis~2 (using information from all four detector cells) for the data taken at the centre of Cell~3 (Position~1, see Figure~\ref{fig:Grid_3pos}). 
    (b) Comparison of Analysis~1 for Position~2 and Position~3 (located at the same distance from the cell centre, see Figure~\ref{fig:Grid_3pos}). No data was recorded at incident angles $\theta_X=0\degree$,~$\theta_Y=0\degree$ and $\theta_X=90\degree$,~$\theta_Y=0\degree$ for Position~3.}
    \label{fig:perc_corr_3pos}
\end{figure}

The resulting fraction of events for which the likelihood-based method correctly identified the incident angle is shown in Figure~\ref{fig:perc_corr_3pos} for the three different positions in Analysis~1, and for Position~1 in Analysis~2. 

The reconstruction generally performs best at large inclinations: A~longer path through the liquid scintillator results in a higher integrated light yield. At very large incident angles in $\theta_Y$, both Cell~3 and Cell~4 will furthermore be crossed, allowing to exclude smaller inclinations due to their detection of significant light yield. This becomes most obvious in the comparison of the performance of Analysis~1 for Positions~2 and~3 at $\theta_Y=75\degree$ in Figure~\ref{fig:perc_corr_3pos_b}, where the particle track crosses Cell~4 in the measurement at Position~2, but not at Position~3. As both positions are symmetric with respect to the centre of the cell and the location of the WOMs, Analysis~1 otherwise exhibits similar performance. 
Analysis~2 provides a higher fraction of correctly assigned incident angles than Analysis~1 (see Figure~\ref{fig:perc_corr_3pos_a}), as was expected from taking into account the information of all four detector cells. This~effect is most evident for small inclinations in $\theta_X$ between $\theta_X=15\degree$ and $\theta_X=45\degree$ with no track or signal in Cell~1. 

\begin{figure}[hbt]
    \centering
    \begin{subfigure}[c]{0.46\textwidth}
        \includegraphics[height=6.6cm]{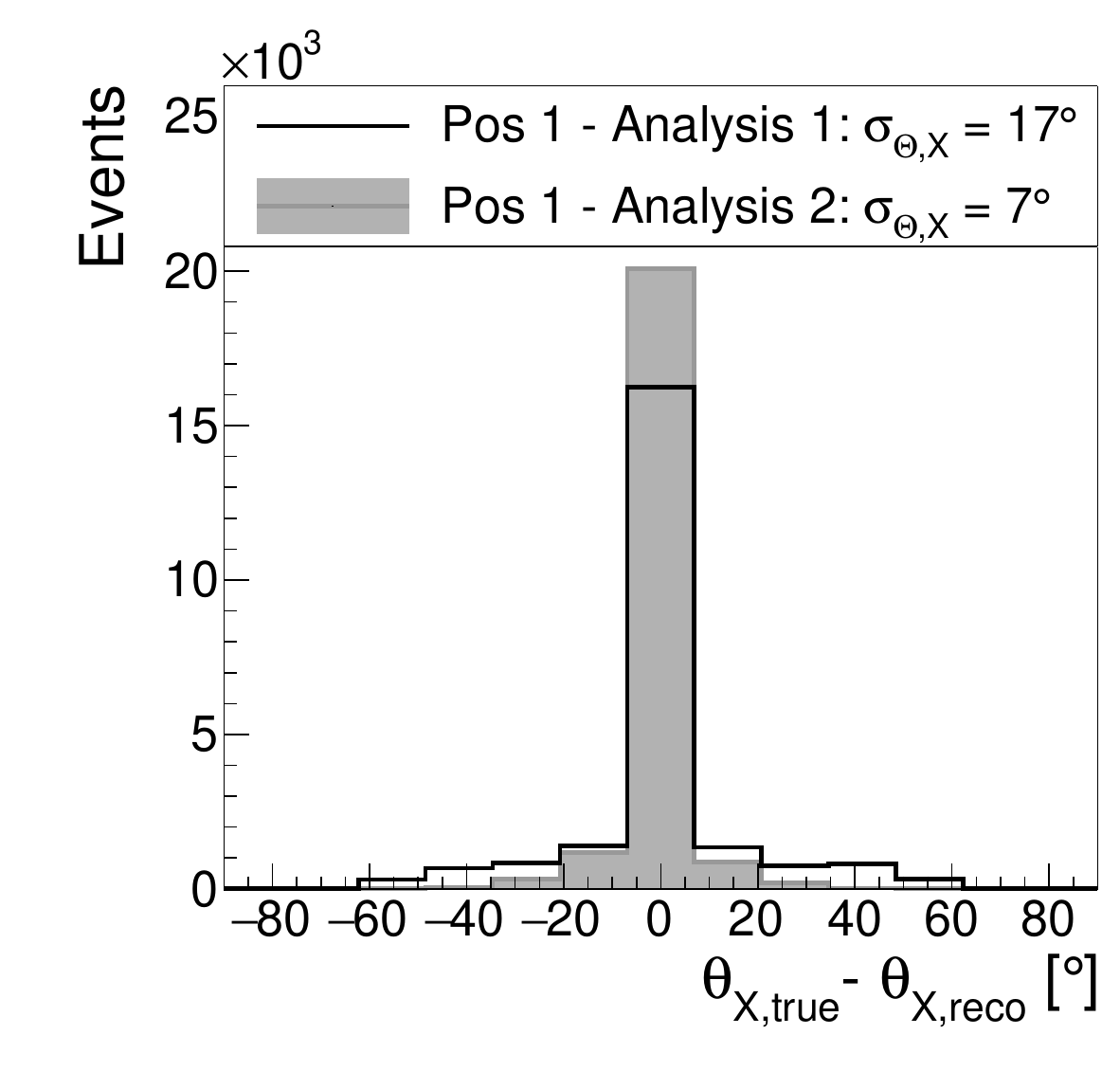}
        \caption{$\theta_X$, Analyses~1 and~2 for Position~1.}
        \label{fig:Res_angle_3pos_a}
    \end{subfigure}   
    \hspace{5.0mm}
    \begin{subfigure}[c]{0.46\textwidth}  
        \includegraphics[height=6.6cm]{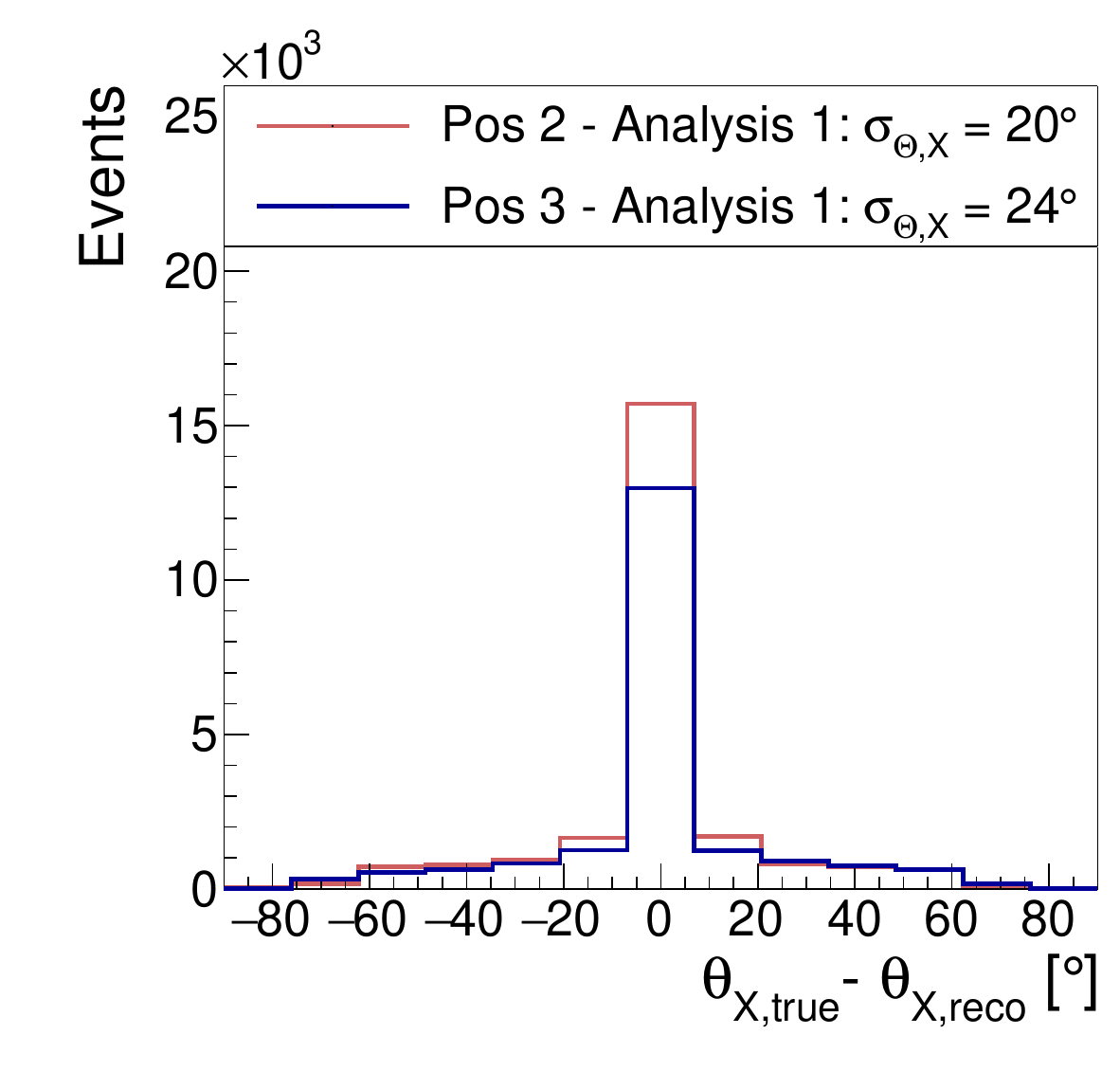}
        \caption{$\theta_X$, Analysis~1 for Positions~2 and~3.}
        \label{fig:Res_angle_3pos_b}
    \end{subfigure}\\
    \vspace{5.0mm}
    \begin{subfigure}[c]{0.46\textwidth}
        \includegraphics[height=6.6cm]{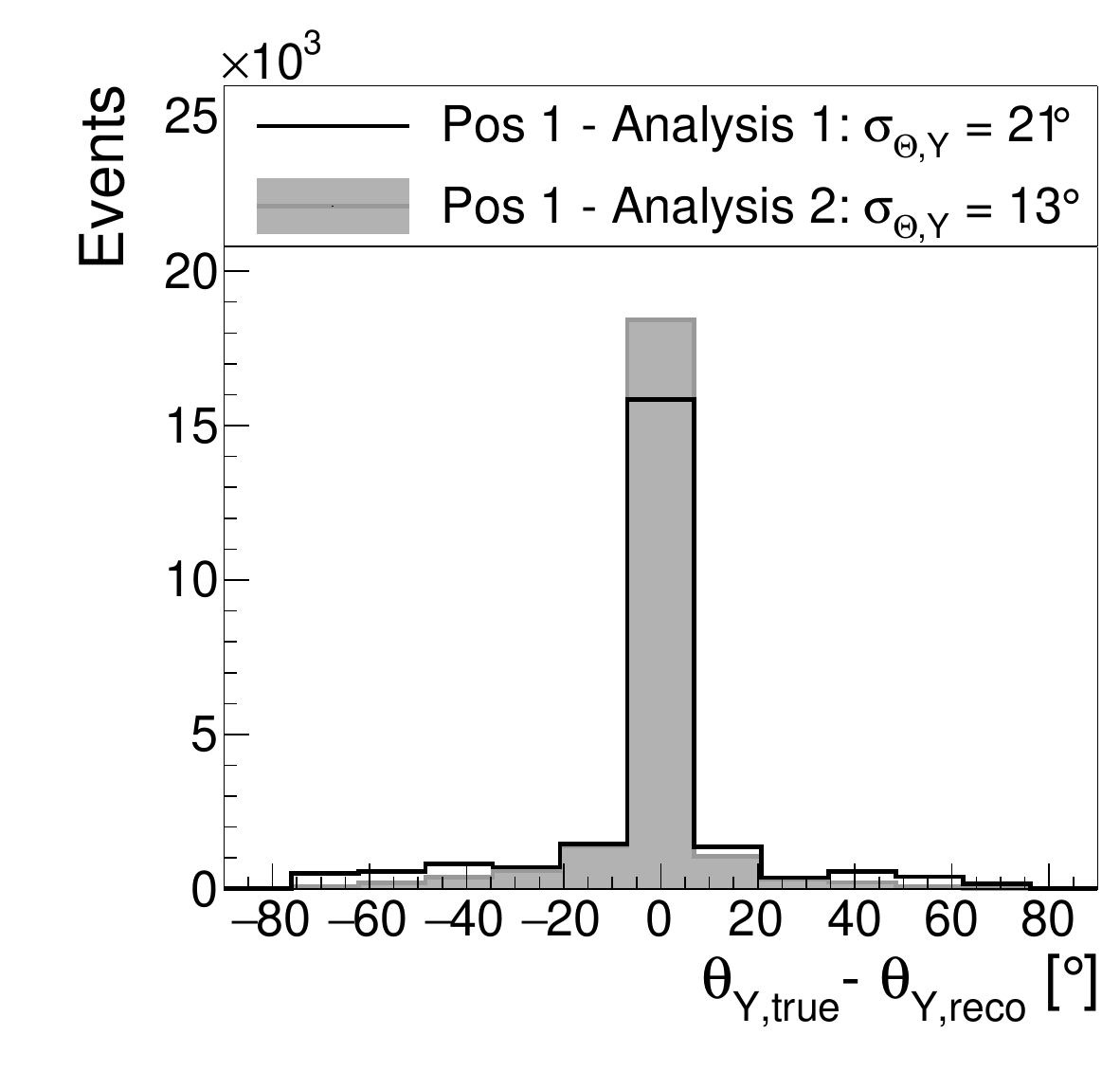}
        \caption{$\theta_Y$, Analyses~1 and~2 for Position~1.}
        \label{fig:Res_angle_3pos_c}
    \end{subfigure}   
    \hspace{5.0mm}
    \begin{subfigure}[c]{0.46\textwidth}
       \includegraphics[height=6.6cm]{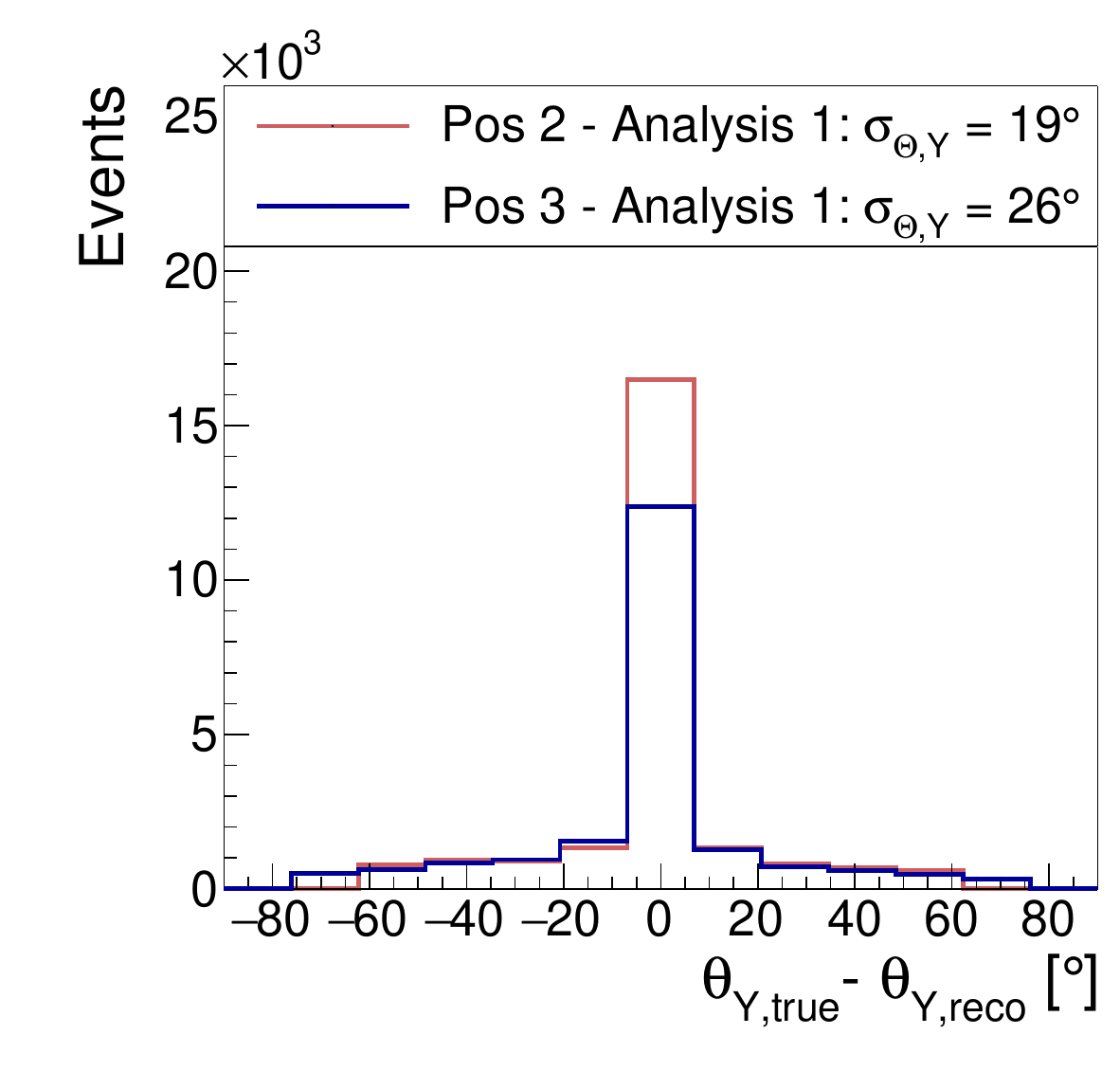}
       \caption{$\theta_Y$, Analysis~1 for Positions~2 and~3.}
       \label{fig:Res_angle_3pos_d}
    \end{subfigure}
    \caption{Difference $\theta_\text{true} - \theta_\text{reco}$ between true and reconstructed incident angles for rotations in~$\theta_X$ and~$\theta_Y$ using the likelihood-based method. (a) and (c) Comparison of Analysis~1 (using only information from Cell~3 and Cell~4) and Analysis~2 (using information from all four detector cells) for the data taken at the centre of~Cell~3 (Position~1, see Figure~\ref{fig:Grid_3pos}). (b) and (d) Comparison of Analysis~1 for Position~2 and Position~3 (located at the same distance from the cell centre, see Figure~\ref{fig:Grid_3pos}).}
    \label{fig:Res_angle_3pos}
\end{figure}

The angular resolution attained by the likelihood-based reconstruction can be quantified by the difference between the true and reconstructed incident angles $\theta_\text{true}$ and $\theta_\text{reco}$. This is shown for rotations in $\theta_X$ around the horizontal axis in Figure~\ref{fig:Res_angle_3pos_a} (Analyses~1 and~2 for Position~1) and Figure~\ref{fig:Res_angle_3pos_b} (Analysis~1 for Positions~2 and~3), and for rotations in $\theta_Y$ around the vertical axis in Figures~\ref{fig:Res_angle_3pos_c}~and~\ref{fig:Res_angle_3pos_d}. At~Position~1, the angular resolutions (given by one standard deviation $\sigma$), attained in Analysis~1 are $\pm17\degree$ for rotations in $\theta_X$ and~$\pm21\degree$ in $\theta_Y$. Using the information of all four detector cells, Analysis~2 provides a significant improvement, resulting in resolutions of $\pm7\degree$~in~$\theta_X$ and $\pm13\degree$~in~$\theta_Y$.

\FloatBarrier
\subsubsection{Particle crossing coordinates and incident angles}
\label{Sec:ReconstructionPointsAngles}
\begin{figure}[hbt]
	\centering
	\includegraphics[width=0.85\textwidth]{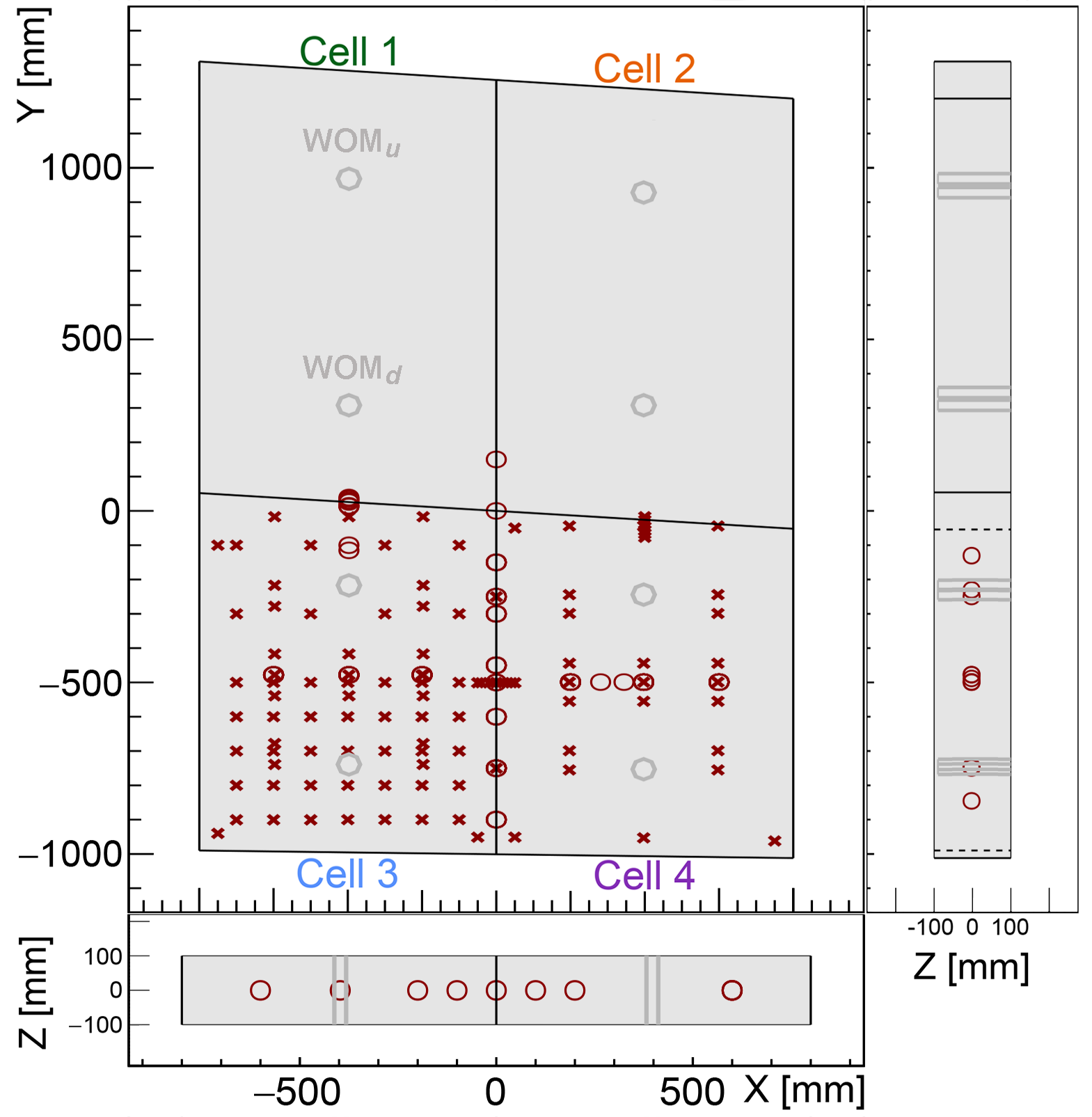}
    \caption{Locations of particle crossing points on the central plane of the detector that were used in the likelihood-based reconstruction of track location and angle. Crosses mark positions that were measured at incident angles of $\theta_X=0\degree$ and $\theta_Y=0\degree$, circles indicate additional measurements where at least one of the two rotation angles was non-zero. The $YZ$ projection on the right shows measurements conducted with the particle beam crossing the detector from the side ($\theta_X=0\degree$, $\theta_Y=90\degree$), the $XZ$ projection at the bottom shows measurements conducted with the particle beam crossing the detector from the bottom ($\theta_X=90\degree$, $\theta_Y=0\degree$).}	
	\label{fig:Grid_249position}
\end{figure}

The likelihood-based reconstruction method (see Section~\ref{Sec:ReconstructionAngles}) was subsequently applied to an even more diverse set of track positions and incident angles on the detector, as illustrated in Figure~\ref{fig:Grid_249position}. 

\begin{figure}[hbt]
    \centering
    \begin{subfigure}[c]{0.46\textwidth}
        \includegraphics[height=6.6cm]{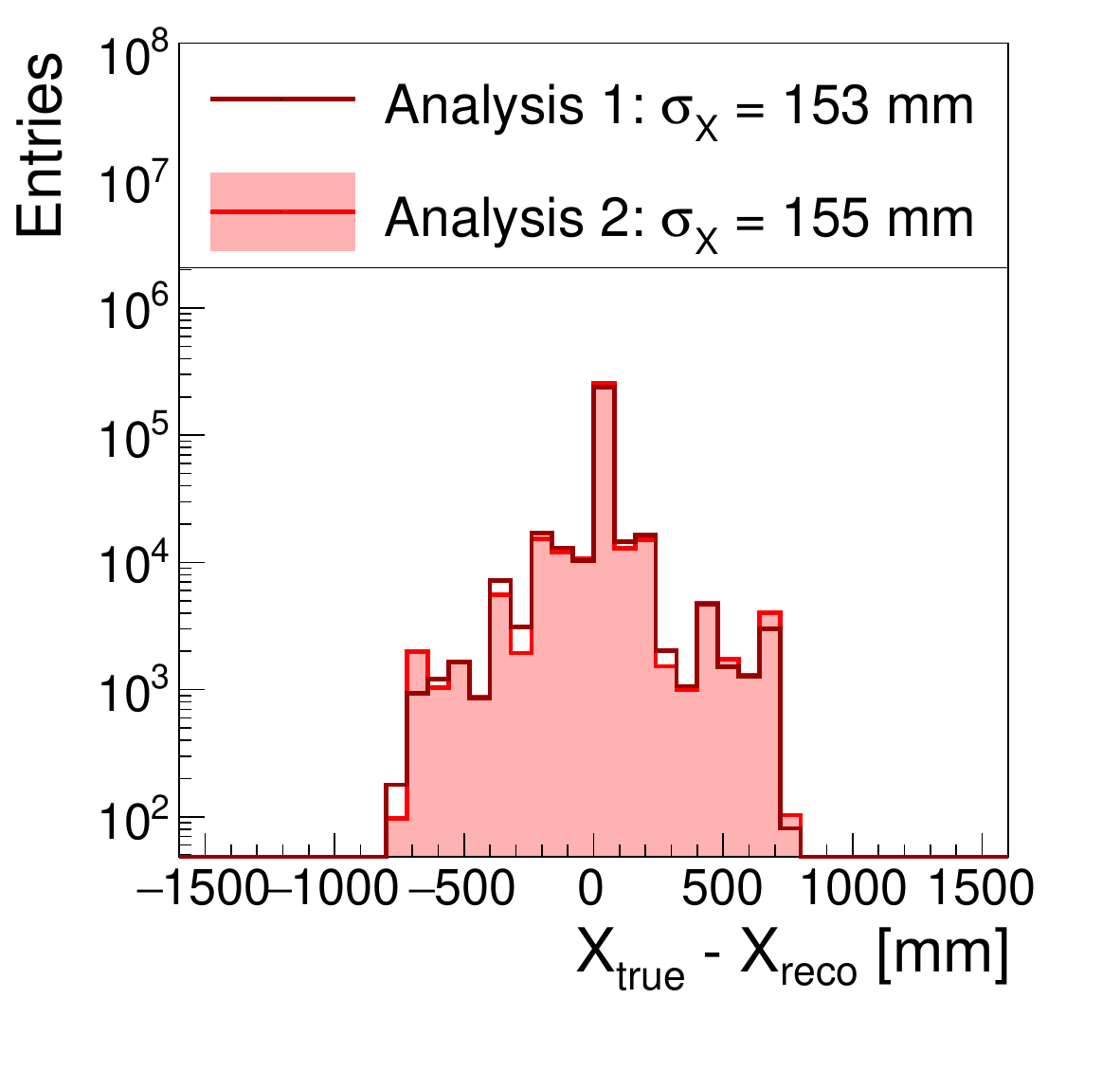}
        \caption{Spatial resolution in $X$.}
        \label{fig:Resolution_XY_249positions_a}
    \end{subfigure}   
    \hspace{5.0mm}
    \begin{subfigure}[c]{0.46\textwidth}   
        \includegraphics[height=6.6cm]{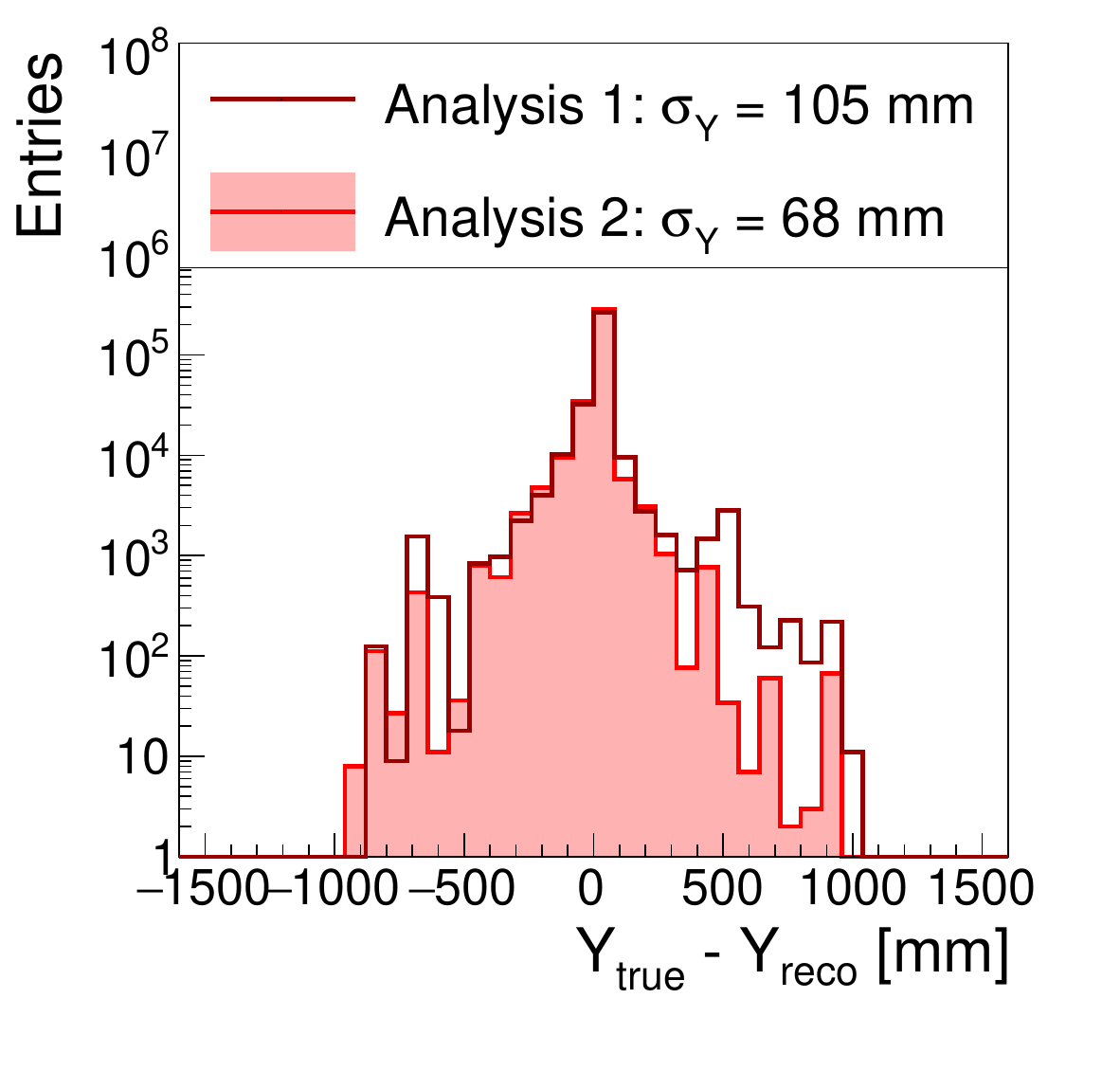}
        \caption{Spatial resolution in $Y$.}
        \label{fig:Resolution_XY_249positions_b}
    \end{subfigure}\\
    \vspace{5.0mm}
    \begin{subfigure}[c]{0.46\textwidth}
        \includegraphics[height=6.6cm]{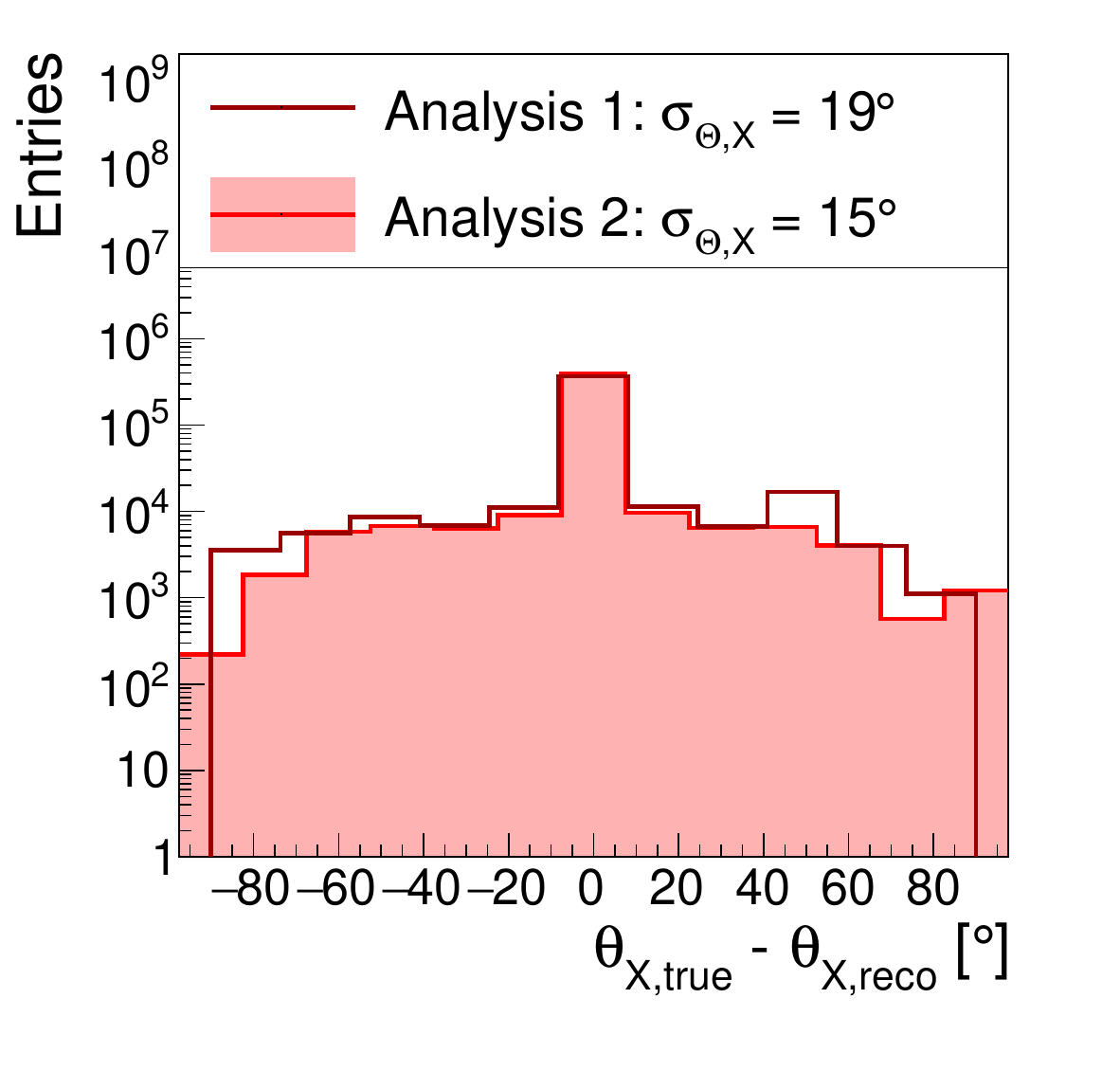}
        \caption{Angular resolution in $\theta_X$.}
        \label{fig:Resolution_angles_249positions_a}
    \end{subfigure}   
    \hspace{5.0mm}
    \begin{subfigure}[c]{0.46\textwidth}  
        \includegraphics[height=6.6cm]{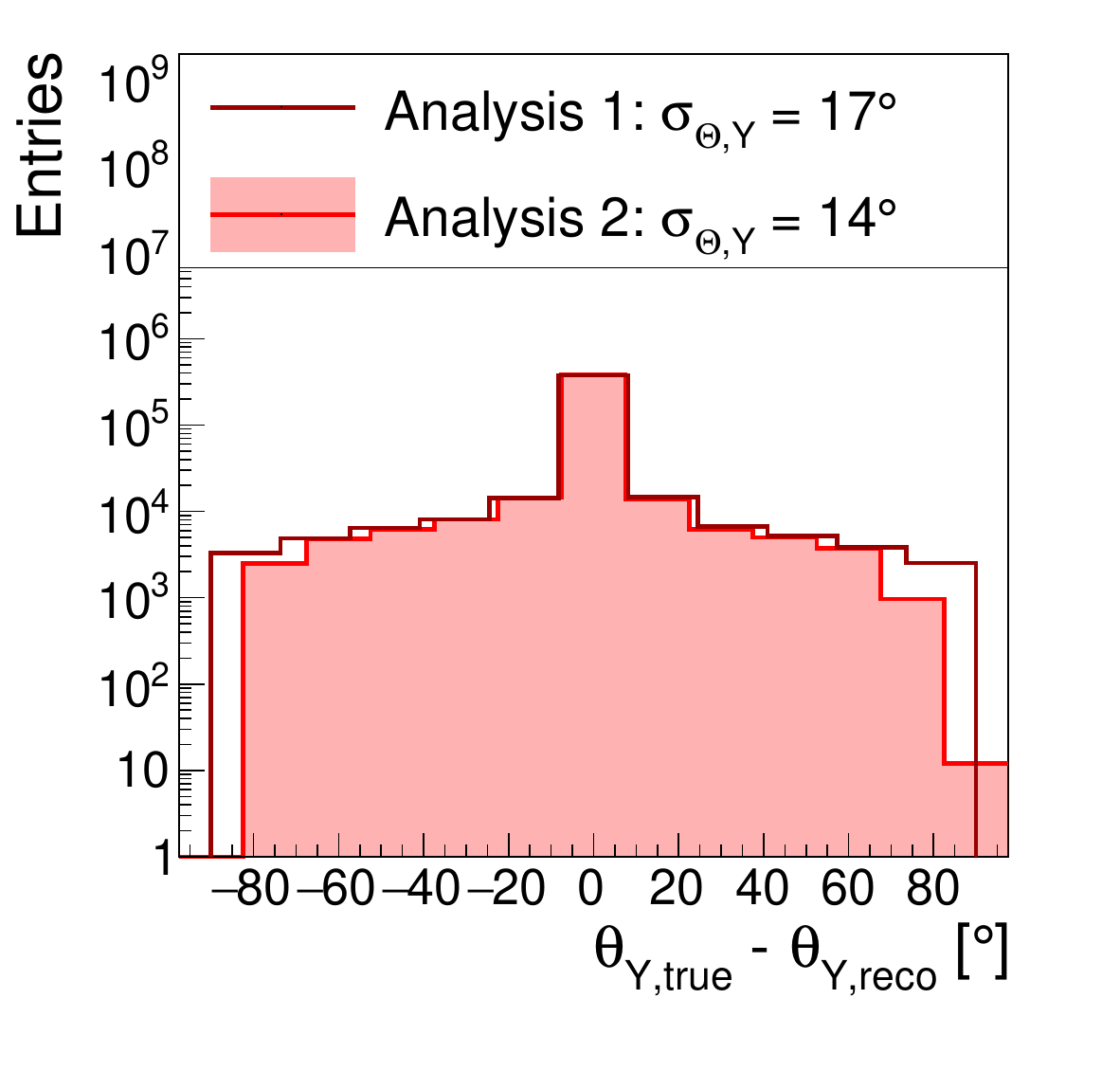}
        \caption{Angular resolution in $\theta_Y$.}
        \label{fig:Resolution_angles_249positions_b}
    \end{subfigure}
    \caption{Differences between true and reconstructed values of particle track coordinates (at~$Z=0$) and inclination attained with the likelihood-based reconstruction for the beam positions shown in Figure~\ref{fig:Grid_249position}. Incident angles $\theta_X$ and~$\theta_Y$ were varied in steps of $15\degree$. Analysis~1 used only information from Cells~3 and~4, Analysis~2 used information from all four detector cells. (a) and (b)~Spatial resolution in $X$ and $Y$. (c) and (d)~Angular resolution in $\theta_X$ and~$\theta_Y$.}
    \label{fig:Res_angle_249pos}
\end{figure}

These analyses used further observables extracted from the data that contain information on the inclination of the particle track: The difference in signal arrival time and the integrated light yield ratio between WOM$_u$ and WOM$_d$ was calculated in each individual detector cell, but also between any pair of WOM$_u$ and WOM$_d$ in two cells adjacent to each other in $X$. Results are shown in Figure~\ref{fig:Res_angle_249pos}, again comparing Analysis~1 (using only information from Cells~3 and~4) and Analysis~2 (using information from all four detector cells).

The differences between true and reconstructed particle track coordinates are shown in Figures~\ref{fig:Resolution_XY_249positions_a}~($X_\text{true} - X_\text{reco}$) and~\ref{fig:Resolution_XY_249positions_b}~($Y_\text{true} - Y_\text{reco}$) for Analyses~1 and~2, along with their standard deviations ($\sigma_X$, $\sigma_Y$) as the attained spatial resolutions. While the standard deviation of the distributions in horizontal direction is essentially the same for both analyses \hbox{($\sigma_{X,1} = 15.3$\,cm and $\sigma_{X,2} = 15.5$\,cm)}, Analysis~2 surpasses Analysis~1 in vertical resolution \hbox{($\sigma_{Y,1} = 10.5$\,cm vs. $\sigma_{X,2} = 6.8$\,cm)} due to the additional information provided by Cells~1 and~2. Both analyses exhibit a significantly better reconstruction in $Y$ than in $X$, which is expected due to the vertical arrangement of WOMs in a SBT detector cell. Figures~\ref{fig:Resolution_angles_249positions_a}~($\theta_{X,\text{true}} - \theta_{X,\text{reco}}$) and~\ref{fig:Resolution_angles_249positions_b}~($\theta_{Y,\text{true}} - \theta_{Y,\text{reco}}$) show the differences between true and reconstructed incidence angles for Analyses~1 and~2. The resulting standard deviations are $\sigma_{\theta_X,1}=19\degree$~and~ $\sigma_{\theta_Y,1}=17\degree$ in Analysis~1 and $\sigma_{\theta_X,2}=15\degree$ and $\sigma_{\theta_Y,1}=14\degree$ in Analysis~2, with Analysis~2 again giving a more precise reconstruction. 

The overall better reconstruction performance is thus obtained by Analysis~2, with an average fraction of 0.48 (see Section~\ref{Sec:ReconstructionAngles}) of correctly identified track crossing points and incident angles (compared to a fraction of 0.45 in Analysis~1). In the four-cell SBT detector prototype, a spatial resolution of $\pm15.5$\,cm in $X$ and $\pm6.8$\,cm in $Y$, and an angular resolution of $\pm15\degree$ in both $\theta_X$ and $\theta_Y$ could be achieved.

\FloatBarrier
\section{Summary and outlook}
\label{Sec:Summary}

A multi-cell WOM-based liquid scintillator (LS) detector prototype has been designed, constructed, and characterised with a test beam as part of the research and development programme for the Surrounding Background Tagger (SBT) of the SHiP experiment. The prototype detector consists of four full-size LS-filled cells constructed from COR-TEN$\textsuperscript{\textregistered}$ steel, whose inner surfaces are coated with a reflective layer of BaSO$_4$ paint. Each cell is instrumented with two Wavelength-shifting Optical Modules (WOMs) coupled to SiPM readout arrays. The prototype was exposed to a 5\,GeV muon beam at the CERN PS T9 test beam facility in order to study the detector response for particles traversing one or several cells from various directions. 

The measurements demonstrate reliable response to minimum ionising particles in all detector cells. As expected, the detected light yield depends on the particle crossing point, inclination angle, and track length within a cell. The largest signals are observed for particle trajectories passing the cell in close vicinity to a WOM, while the lowest light yields are measured near the cell corners. For cells of different sizes, differences in the integrated light yield are observed that follow the expectations derived from detector geometry. A likelihood-based correction, using the reconstructed particle crossing coordinates, is applied to compensate for position-dependent variations of the integrated light yield. This method manages to significantly improve the response uniformity within a detector cell while only using observables measured by the detector itself.  
A~similar technique is employed for the timing information: Due to different paths of photon propagation within the LS cells, the average measured signal arrival time calculated from the WOM signals varies across the detector volume. By applying a likelihood-based correction, the timing variations can be substantially reduced to better than 0.4\,ns. 

The measured detector response is compared to a detailed GEANT4-simulation, where good qualitative agreement is obtained for all studied particle crossing positions. The comparison further indicates that the effective reflectivity of the BaSO$_4$-coated inner cell surface is lower than the nominal reflectivity quoted by the manufacturer of the paint: A simulation using $75\%$ of the nominal values provides the best agreement with the measured light yield distributions. Based on this finding, we started to investigate better reflector materials~\cite{Brignoli:2025spc}.

The multi-cell detector prototype allowed to study for the first time particle tracks crossing several SBT cells, which represent an important topology for the final detector. The measurements conducted in this study demonstrate that the total integrated light yield recorded for tracks crossing neighbouring detector cells depend on both the track length inside the LS and the location of the particle trajectory relative to the WOMs. A likelihood-based reconstruction procedure is applied to the combined information of charge and timing from several detector cells, allowing the simultaneous reconstruction of particle crossing positions and incident angles. Its performance improves when information from all detector cells -- also those without signal -- is included. In the analysed data sample, a spatial resolution of $\pm$15.5\,cm in horizontal and $\pm$6.8\,cm in vertical direction is achieved, while the angular resolution is approximately $\pm$15$\degree$ for both inclinations. 

The results of this study demonstrate that a segmented WOM-based liquid-scintillator detector is able to well reconstruct the trajectories of particles crossing several detector cells. The quantitative agreement of measurements and simulation establishes a solid basis for future detector development and optimisation of the reconstruction methods for the SHiP Surrounding Background Tagger.

\FloatBarrier
\section*{Acknowledgements}
The measurements that led to these results were carried out at the CERN East Area test beam facility. We would like to express our sincere gratitude to the CERN management and staff, particularly the test beam coordinators, for their constant support. We would also like to express our gratitude for the skills and efforts of the technicians at our institutions. We thank S.~Bordoni from the University of Geneva for providing the 64-channel WaveCatcher crate that was used in the data acquisition. We thank D.~Breton and J.~Maalmi from IJCLab for their help in upgrading the WaveCatcher with an additional 8~channels, allowing us to record all signals in one crate. We are grateful to MPIK Heidelberg, especially C.~Buck and B.~Gramlich, for providing access to their purification plant facilities and help in purifying the liquid scintillator used in this detector prototype.
We thank J.~Rack-Helleis of JGU Mainz for performing efficiency measurements on our WOM tubes. We acknowledge the fruitful discussions with all the individuals from JGU Mainz and DESY involved in establishing the WOM as a sensor for IceCube. 
We thank the Deutsche Forschungsgemeinschaft~(DFG) for funding support within grant~289921825 and the Bundesministerium für Bildung und Forschung~(BMBF) for funding support within the High-D consortium. The test beam measurements have received funding from the European Union’s Horizon Europe Research and Innovation programme under Grant Agreement No.~101057511 (EURO-LABS).

\FloatBarrier

\end{document}